\documentclass{article}%
\usepackage{amsfonts}
\usepackage{amsmath}
\usepackage{amssymb}
\usepackage{graphicx}%
\usepackage{subcaption}
\providecommand{\U}[1]{\protect\rule{.1in}{.1in}}
\newtheorem{theorem}{Theorem}

\newtheorem{example}[theorem]{Example}

\newtheorem{remark}[theorem]{Remark}

\begin{document}

\title{Option pricing in affine generalized Merton models}
\author{Christian Bayer and John Schoenmakers}
\maketitle

\begin{abstract}
In this article we consider affine generalizations of the Merton jump diffusion model
\cite{Mer} and the respective pricing of European options. On the one hand,
the Brownian motion part in the Merton model may be generalized to a
log-Heston model, and on the other hand, the jump part may be generalized to
an affine process with possibly state dependent jumps. While the
characteristic function of the log-Heston component is known in closed form,
the characteristic function of the second component may be unknown explicitly.
For the latter component we propose an approximation procedure based on the
method introduced in \cite{BKS09}. We conclude with some numerical examples.

\end{abstract}

\section{Introduction}

The Merton jump diffusion model \cite{Mer} can be considered one of the first
asset models beyond Black-Scholes that may produce non-flat implied volatility
surfaces. On the other hand, European options within this model can be priced
quasi-analytically by means of an infinite series of Black-Scholes type
expressions. From a mathematical point of view, the logarithm of the Merton
model is the sum of a compound Poisson process an an independent Brownian
motion, and as such can be seen as the sum of two independent degenerate
affine processes. The goal of this article is to enlarge the flexibility of
the Merton model by generalizing the Brownian motion to a continuous affine
Heston model and replacing the compound Poisson process by another,
independent, affine model that may incorporate both stochastic volatility and
jumps. In financial modeling affine processes have become very popular the
last decades, both due to their flexibility and their analytical tractability.
The theoretical analysis of affine processes is developed in the seminal
papers \cite{DPS} and \cite{DFS}. Once the characteristic functions of the
affine ingredients of our new generalized Merton model are known, we may price
European options by the meanwhile standard Carr-Madan Fourier based method
\cite{CM}. For a variety of affine models, such as the Heston model and
several stochastic volatility models with state independent jumps, the
characteristic function is explicitly known. However if, for instance, in an
affine jump model the jump intensity depends on the present state, a closed
form expression for the characteristic function is not known to the best of
our knowledge. Yet, such models make sense in certain applications such as
crisis modeling. For example, one may wish to model an increased intensity of
downward jumps in regimes of increased volatility. In order to cope with
such kind of processes numerically, we recap and apply the general series expansion representation
for the characteristic function of an affine process developed in \cite{BKS09} and present some
numerical examples.

\section{Merton jump diffusion models}

Merton~\cite{Mer} introduced and studied stock price models of the form%
\[
S_{t}=S_{0}e^{rt+Y_{t}},
\]
where $Y$ is the sum of a Brownian motion with drift and an independent
compound Poisson process,%
\begin{equation}
Y_{t}=\gamma t+\sigma W_{t}+J_{t}, \label{lnx}%
\end{equation}
and $r$ is a constant, continuously compounded risk-free rate. In (\ref{lnx})
$J$ may be represented as
\[
J_{t}=\sum_{l=1}^{N_{t}}U_{l},
\]
where $U_{1},U_{2},...$ are i.i.d. real valued random variables and $N_{t}$
denotes the number of time marks up to time $t$ that arrive at exponential
times with parameter $\lambda,$ i.e.
\[
N_{t}:=\#\left\{  i:s_{i}\leq t,\text{ \ \ }s_{i}-s_{i-1}\sim\exp_{\lambda
},\text{ \ \ }i=1,2,...\right\}
\]
with $s_{0}:=0,$ and where $\tau\sim\exp_{\lambda}$ denotes an exponentially
distributed random variable with
\[
\mathbb{P}\left[  \tau\geq s\right]  =e^{-\lambda s}\text{ for all }s\geq0.
\]
From basic probability theory we know that $N_{t}$ is Poisson distributed
according to%
\[
\mathbb{P}\left[  N_{t}=n\right]  =e^{-\lambda t}\frac{\left(  \lambda
t\right)  ^{n}}{n!},
\]
and that the characteristic function of $Y_{t}$ is given by,%
\begin{align}
\Phi_{t}(z)  &  =\mathbb{E}\left[  e^{\mathfrak{i}zY_{t}}\right]
=e^{\mathfrak{i}z\gamma t}\mathbb{E}\left[  e^{\mathfrak{i}z\sigma W_{t}%
}\right]  \mathbb{E}\left[  e^{\mathfrak{i}zJ_{t}}\right] \nonumber\\
&  =\exp\left[  \mathfrak{i}z\gamma t-\frac{z^{2}\sigma^{2}}{2}t+\lambda
t\int\left(  e^{\mathfrak{i}zu}-1\right)  p(du)\right]  , \label{cfm}%
\end{align}
for a certain jump probability measure $p$ on $\mathcal{B}(\mathbb{R})$ due to
the distribution of $U_{1}.$

We henceforth assume a risk-neutral pricing measure and due to no-arbitrage
arguments we must have that $S_{t}e^{-rt}$ is a martingale under this measure.
This implies that%
\begin{equation}
S_{0}=\mathbb{E}\left[  S_{t}e^{-rt}\right]  =S_{0}\mathbb{E}\left[
\exp\left(  Y_{t}\right)  \right]  =S_{0}\Phi_{t}(-\mathfrak{i}),\text{
\ \ hence \ \ }\Phi_{t}(-\mathfrak{i})=1. \label{mac}%
\end{equation}
By (\ref{cfm}) we then get%
\begin{equation}
\gamma=-\frac{\sigma^{2}}{2}-\lambda\int\left(  e^{u}-1\right)  p(du).
\label{re}%
\end{equation}
As an example, with $\lambda=0$ (no jumps), $\gamma=-\frac{\sigma^{2}}{2}$ and
we retrieve the risk neutral Black-Scholes model. Merton particularly studied
the case where $U$ is normally distributed and derived a representation for a
call (or put) option in terms of an infinite series of Black-Scholes
expressions. In this paper we are interested in generalizations of (\ref{lnx})
of the form%
\begin{equation}
Y_{t}=\gamma t+\sigma W_{t}+X_{t}^{1}, \label{glnx}%
\end{equation}
or even,
\begin{equation}
Y_{t}=\gamma t+H_{t}+X_{t}^{1}, \label{glnx1}%
\end{equation}
where $H$ is the first component of a log-Heston type model with $H_{0}=0,$
whereas $X_{t}^{1}$ is the first component of some generally multidimensional
affine (eventually jump) process $X,$ independent of $W$ and $H$ respectively,
with $X_{0}^{1}=0.$ In particular, the characteristic function of $X^{1}$ is
possibly not known in closed form.

\section{Recap of affine processes and approximate characteristic functions}

We consider an affine process $X$ in the state space $\mathfrak{X}%
\subset\mathbb{R}^{d},$ $d\in\mathbb{N}_{+},$ with generator given by%
\begin{align}
Af(x)  &  =\frac{1}{2}\sum_{i,j=1}^{d}a^{ij}(x)\frac{\partial^{2}f}{\partial
x^{i}\partial x^{j}}+\sum_{i=1}^{d}b^{i}(x)\frac{\partial f}{\partial x^{i}%
}\label{Gen}\\
&  +\int_{\mathbb{R}^{d}}\left[  f(x+z)-f(x)-z^{\top}\frac{\partial
f}{\partial x}\right]  v(x,dz),\nonumber
\end{align}
where $a^{ij}$ and $b^{i}$ are suitably defined \emph{affine} functions in $x$
on $\mathbb{R}^{d},$ and%
\[
v(x,dz)=:v^{0}(dz)+x^{\top}v^{1}(dz)
\]
with $v^{0}$ and $v_{i}^{1},$ \thinspace$i=1,...,d,$ being suitably defined
locally finite measures on $\mathcal{B}(\mathbb{R}^{d}\backslash\left\{
0\right\}  ).$ Alternatively, the dynamics of $X$ are described by the
It\^{o}-L\'{e}vy SDE:
\begin{equation}
dX_{t}=b(X_{t})dt+\sigma(X_{t})dW(t)+\int_{\mathbb{R}^{d}}z\widetilde
{N}(X_{t-},dt,dz),\quad X_{0}=x, \label{ItoLev}%
\end{equation}
where $W$ is a Wiener process in $\mathbb{R}^{m}$ and the function
$\sigma:\mathbb{R}^{d}\rightarrow\mathbb{R}^{d}\times\mathbb{R}^{m}$
satisfies
\[
\sum_{k=1}^{m}\sigma_{ik}(x)\sigma_{jk}(x)=a^{ij}(x).
\]
Further, in (\ref{ItoLev})
\[
\widetilde{N}(x,dt,dz):=\widetilde{N}(x,dt,dz,\omega):=N(x,dt,dz,\omega
)-v(x,dz)dt,
\]
is a compensated Poisson point process on $\mathbb{R}_{+}\times\mathbb{R}%
^{d},$ such that%
\[
\mathbb{P}\left[  N(x,(0,t],B)=k\right]  =\exp(-tv(x,B))\frac{t^{k}v^{k}%
(x,B)}{k!},\quad k=0,1,2,...
\]
for bounded $B\in\mathcal{B}(\mathbb{R}^{d}\backslash\left\{  0\right\}  ).$
It is assumed that the coefficients in (\ref{ItoLev}) (and so in (\ref{Gen}))
satisfy sufficient conditions such that (\ref{ItoLev}) has a unique strong
solution $X,$ and that $X$ is an affine process with generator (\ref{Gen}).
For details regarding these assumptions, in particular the admissibility
conditions that are to be fulfilled, we refer to \cite{BKS09}, \cite{DFS}, see
also \cite{DPS}.

The characteristic function of $X_{t}^{0;x},$ with $X_{0}^{0;x}=x\in
\mathbb{R}^{d},$ is denoted by,%
\begin{equation}
\widehat{p}(t,x,u):=\mathbb{E}\left[  e^{\mathfrak{i}u^{\top}X_{t}^{0;x}%
}\right]  ,\text{ \ \ }x\in\mathfrak{X},\text{ }u\in\mathbb{R}^{d},\text{
\ \ }t\geq0. \label{char}%
\end{equation}

For a variety of affine processes the characteristic function is explicitly
known. However, in general the characteristic function of an affine process
involves the solution of a multi-dimensional Riccati equation that may not be
solved explicitly. In particular, for affine jump processes with state
dependent jump part a closed form expression for the characteristic function
generally doesn't exist. In this section we recall the approach by Belomestny,
Kampen, and Schoenmakers \cite{BKS09}, who developed in general a series
expansion for the log-characteristic function in terms of the ingredients of
the generator of the affine process under consideration. By truncating this
expansion one may obtain an approximation of the characteristic function that
may subsequently be used for approximate option pricing.

Henceforth, $x\in\mathfrak{X}$ is fixed. It is assumed that the characteristic
function (\ref{char}) satisfies:\newline\ \newline Assumption \textbf{HE:}
There exists a non-increasing function $R:(0,\infty)\ni r\rightarrow
R(r)\in(0,\infty],$ such that for any $u\in\mathbb{R}^{d},$ the function
$[0,\infty)\ni s\rightarrow\widehat{p}(s,x,u)\in\mathbb{C}$ has a holomorphic
extension to the region%
\[
G_{u}:=\left\{  z\in\mathbb{C}:\left\vert z\right\vert <R\left(  \left\Vert
u\right\Vert \right)  \right\}  \cup\left\{  z\in\mathbb{C}:\operatorname{Re}%
z\geq0\text{ \ \ and \ \ }\left\vert \operatorname{Im}z\right\vert <R\left(
\left\Vert u\right\Vert \right)  \right\}
\]
(cf. Prop. 3.7, 3.8, and Th. 4.1 and Corr. 4.2-4.4 in \cite{BKS09}).

Under Assumption \textbf{HE}, Th. 3.4 in \cite{BKS09} is particularly
fulfilled for each $u.$ Moreover, by taking in \cite{BKS09}, Th. 3.4-(ii),%
\begin{equation}
\eta_{u}=\eta(\left\Vert u\right\Vert ):=\frac{\pi}{2R\left(  \left\Vert
u\right\Vert \right)  }, \label{et}%
\end{equation}
we arrive at the log-series representation \cite{BKS09}-(5.12) for the
characteristic function,%
\begin{align}
\ln\widehat{p}(t,x,u)  &  =\ln\left(  \sum_{r\geq0}h_{r,0}(u;\eta
_{u})(1-e^{-\eta_{u}t})^{r}\right)  +\mathfrak{i}u^{\top}x\label{logseries}\\
&  +x^{\top}\frac{\sum_{r\geq1}h_{r}(u;\eta_{u})\,(1-e^{-\eta_{u}t})^{r}}%
{\sum_{r\geq0}h_{r,0}(u;\eta_{u})(1-e^{-\eta_{u}t})^{r}}\mathfrak{,}\text{
\ \ }u\in\mathbb{R}^{d},\text{ \ \ }t\geq0,\nonumber
\end{align}
where the coefficients $h_{r,0}(u;\eta_{u})\in\mathbb{C}$ and $h_{r}%
(u;\eta_{u})=\left[  h_{r,e_{1}}(u;\eta_{u}),...,h_{r,e_{d}}(u;\eta
_{u})\right]  \in\mathbb{C}^{d}$ with $e_{i}:=\left(  \delta_{ij}\right)
_{j=1,...,d},$ can be computed algebraically from the coefficients of the
affine generator $A$ in a way that is described below.

Alternatively, in \cite{BKS09} a ground expansion of the form%
\begin{equation}
\widehat{p}(t,x,u)=e^{\mathfrak{i}u^{\top}x}\sum_{r=0}^{\infty}q_{r}%
(x,u;\eta_{u})(1-e^{-\eta_{u}t})^{r}\mathfrak{,}\text{ \ \ }u\in\mathbb{R}%
^{d},\text{ \ \ }t\geq0, \label{boden}%
\end{equation}
is derived with%
\[
q_{r}(x,u;\eta_{u})=\sum_{\left\vert \gamma\right\vert \leq r}h_{r,\gamma
}(u;\eta_{u})x^{\gamma},
\]
and the $h_{r,\gamma}$ are computed by the recursion (\ref{Rec}) as described below.

\begin{remark}
Because of Assumption \textbf{HE,} if Th. 3.4-(i) applies for some $u,$ it
applies for any $u^{\prime}$ with $\left\Vert u^{\prime}\right\Vert
\leq\left\Vert u\right\Vert $ $.$ As a consequence, one may take in
(\ref{logseries}) any $\eta_{u}=\eta(\left\Vert u^{\prime\prime}\right\Vert )$
with $\left\Vert u^{\prime\prime}\right\Vert \geq\left\Vert u\right\Vert .$
\end{remark}

In order to outline the construction of the expansion (\ref{logseries}), let
us denote%
\begin{equation}
f_{u}(x):=e^{\mathfrak{i}u^{\top}x},\quad z\in\mathbb{R}^{d}. \label{SF}%
\end{equation}
Then for each multi-index $\beta\in$ $\mathbb{N}_{0}^{d}$ we may compute
algebraically%
\begin{equation}
\mathfrak{b}_{\beta}(x,u):=\mathfrak{i}^{-\left\vert \beta\right\vert
}\partial_{u^{\beta}}\frac{Af_{u}(x)}{f_{u}(x)}=:\mathfrak{b}_{\beta}%
^{0}(u)+\sum_{\kappa,\,\left\vert \kappa\right\vert =1}\mathfrak{b}%
_{\beta,\kappa}^{1}(u)\,x^{\kappa} \label{sym}%
\end{equation}
(in multi-index notation), provided that for the jump part in the generator
(\ref{Gen}),%
\begin{align*}
&  \frac{1}{f_{u}(x)}\int_{\mathbb{R}^{d}}\left(  f_{u}(x+z)-f_{u}(x)-z^{\top
}\frac{\partial f_{u}}{\partial x}\right)  v(x,dz)\\
&  =\int_{\mathbb{R}^{d}}\left(  e^{\mathfrak{i}u^{\top}z}-1-\mathfrak{i}%
u^{\top}z\right)  v^{0}(dz)+x^{\top}\int_{\mathbb{R}^{d}}\left(
e^{\mathfrak{i}u^{\top}z}-1-\mathfrak{i}u^{\top}z\right)  v^{1}(dz)
\end{align*}
is explicitly known. That is, the cumulant generating functions of $v^{0}$ and
$v_{i}^{1},$ $i=1,...,d,$ are explicitly known. We note that the expression
$Af_{u}(x)/f_{u}(x)$ in (\ref{sym}) is termed the \textit{symbol} of the
operator $A.$ As such the $\mathfrak{b}_{\beta}$ in (\ref{sym}) are, modulo
some integer power of the imaginary unit, derivatives of the symbol of $A.$

Let us next consider a fixed $u\in\mathbb{R}^{d}$ and $\eta_{u}>0.$ Then for
each multi-index $\gamma$ and integer $r\geq0$ we are going to construct
$h_{r,\gamma}=h_{r,\gamma}(u;\eta_{u})$ as follows. For $\left\vert
\gamma\right\vert >r$ we set $h_{r,\gamma}\equiv0$ and for $0\leq r\leq$
$\left\vert \gamma\right\vert ,$ the $h_{r,\gamma}$ are determined by the
following recursion. As initialization we take $h_{0,0}\equiv1,$ and for
$0\leq r<$ $\left\vert \gamma\right\vert $ we have (cf. \cite{BKS09}-(4.6)),
\begin{align}
(r+1)h_{r+1,\gamma}  &  =\sum_{\left\vert \beta\right\vert \leq r-\left\vert
\gamma\right\vert }\eta_{u}^{-1}\binom{\gamma+\beta}{\beta}h_{r,\gamma+\beta
}\mathfrak{b}_{\beta}^{0}\label{Rec}\\
&  +\!\!\!\!\!\!\sum_{\left\vert \kappa\right\vert =1,\,\kappa\leq\gamma
}\ \sum_{\left\vert \beta\right\vert \leq r+1-\left\vert \gamma\right\vert
}\!\!\!\eta_{u}^{-1}\binom{\gamma-\kappa+\beta}{\beta}h_{r,\gamma-\kappa
+\beta}\mathfrak{b}_{\beta,\kappa}^{1}+rh_{r,\gamma},\nonumber
\end{align}
where $\left\vert \gamma\right\vert \leq r+1,$ and empty sums are defined to
be zero. We next set%
\[
h_{r}(u;\eta_{u}):=\left[  h_{r,e_{i}}(u;\eta_{u})\right]  _{i=1,...,d}.
\]
In view of Th. 4.1 in \cite{BKS09} suitable choices of $\eta_{u}$ are%
\begin{align*}
\eta_{u}  &  \gtrsim1+\left\Vert u\right\Vert ^{2}\text{ \ \ in case of pure
affine diffusions,}\\
\eta_{u}  &  \gtrsim e^{\zeta\left\Vert u\right\Vert },\text{ \ }%
\zeta>0,\text{ for affine jump processes with thinly tailed large jumps.}%
\end{align*}
In practice the best choice of $\eta_{u}$ can be determined in view of the
particular problem under consideration. Generally, on the one hand, $\eta_{u}$
should be large enough to guarantee convergence of the series (\ref{logseries}%
), but on the other hand should not taken to be unnecessarily large for this
would result in series that converges too slowly.

As a natural approximation to (\ref{logseries}) and (\ref{boden}) we consider
for $K=1,2,...,$
\begin{gather}
\ln\widehat{p}_{K}(t,x,u)=\ln\left(  \sum_{r=0}^{K}h_{r,0}(u;\eta
_{u})(1-e^{-\eta_{u}t})^{r}\right)  +\mathfrak{i}u^{\top}x\label{appr}\\
+x^{\top}\frac{\sum_{r=1}^{K}h_{r}(u;\eta_{u})\,(1-e^{-\eta_{u}t})^{r}}%
{\sum_{r=0}^{K}h_{r,0}(u;\eta_{u})(1-e^{-\eta_{u}t})^{r}}\mathfrak{,}\text{
\ \ }u\in\mathbb{R}^{d},\text{ \ \ }t\geq0,\nonumber
\end{gather}
and the ground expansion based approximation
\begin{equation}
\widehat{p}(t,x,u)=e^{\mathfrak{i}u^{\top}x}\sum_{r=0}^{K}q_{r}(x,u;\eta
_{u})(1-e^{-\eta_{u}t})^{r}\mathfrak{,}\text{ \ \ }u\in\mathbb{R}^{d},\text{
\ \ }t\geq0, \label{appr1}%
\end{equation}
respectively.

\begin{remark}
\label{rem:heuristic-eta}
In connection with approximations (\ref{appr}) and (\ref{appr1}) it seems
natural to estimate $R_{u}$ in view of Cauchy's criterion, and $\eta_{u}$
according to (\ref{et}). That is, we could take%
\[
\eta_{u}\approx\frac{\pi}{2}\sqrt[K]{\frac{\left\vert A^{K}f_{u}(x)\right\vert
}{K!}},
\]
where the sequence $g_{r}(x,u):=A^{r}f_{u}(x)/f_{u}(x)$ can be obtained from
the recursion%
\begin{align}
g_{r+1,\gamma} &  =\sum_{\left\vert \beta\right\vert \leq r-\left\vert
\gamma\right\vert }\binom{\gamma+\beta}{\beta}g_{r,\gamma+\beta}%
\mathfrak{b}_{\beta}^{0}\label{grecu}\\
&  +\sum_{\left\vert \kappa\right\vert =1,\,\kappa\leq\gamma}\ \sum
_{\left\vert \beta\right\vert \leq r+1-\left\vert \gamma\right\vert }%
\binom{\gamma-\kappa+\beta}{\beta}g_{r,\gamma-\kappa+\beta}\mathfrak{b}%
_{\beta,\kappa}^{1},\nonumber
\end{align}
with $g_{0,0}=1$ (cf \cite{BKS09}-(4.6)).
\end{remark}

\section{Generalized Merton models}

We now consider generalized Merton models of the form (\ref{glnx}) and
(\ref{glnx1}). For the characteristic function of (\ref{glnx}) we have,
\begin{align}
\Phi_{t}(z)  &  =e^{\mathfrak{i}z\gamma t}\mathbb{E}e^{\mathfrak{i}z\sigma
W_{t}}\mathbb{E}e^{\mathfrak{i}zX_{t}^{0;\left(  0,x^{2},...,x^{d}\right)
;1}}\nonumber\\
&  =\exp\left[  \mathfrak{i}z\gamma t-\frac{z^{2}\sigma^{2}}{2}t\right]
\widehat{p}(t,\left(  0,x^{2},...,x^{d}\right)  ,\left(  z,0,...,0\right)  ),
\label{gcfm}%
\end{align}
where $X_{t}^{\cdot\cdot\cdot;1}$ denotes the first component of $X_{t}%
^{\cdot\cdot\cdot}$ cf. (\ref{cfm}). Firstly, the martingale condition
(\ref{mac}) can now be formulated as%
\begin{equation}
\gamma=-\frac{\sigma^{2}}{2}-t^{-1}\ln\widehat{p}(t,\left(  0,x^{2}%
,...,x^{d}\right)  ,\left(  -\mathfrak{i},0,...,0\right)  ), \label{gam}%
\end{equation}
that\bigskip\ is, $\gamma$ may in principle depend on time $t.$ More
generally, the characteristic function of (\ref{glnx1}) takes the form,%
\begin{equation}
\Phi_{t}(z)=e^{\mathfrak{i}z\gamma t}\widehat{p}_{H}(t,z)\widehat{p}(t,\left(
0,x^{2},...,x^{d}\right)  ,\left(  z,0,...,0\right)  ), \label{gcfm1}%
\end{equation}
with $\widehat{p}_{H}(t,z):=\mathbb{E}\left[  \exp(\mathfrak{i}zH_{t})\right]
,$ and%
\begin{equation}
\gamma=-t^{-1}\ln\widehat{p}_{H}(t,-\mathfrak{i})-t^{-1}\ln\widehat
{p}(t,\left(  0,x^{2},...,x^{d}\right)  ,\left(  -\mathfrak{i},0,...,0\right)
). \label{gam1}%
\end{equation}
In a situation where $\widehat{p}$ in (\ref{gcfm}) and (\ref{gcfm1}),
respectively, is unknown in closed form, we propose to replace it with an
approximation $\widehat{p}_{K}$ due to (\ref{appr}) for some level $K$ large
enough. It is convenient to choose $X_{t}^{1}$ and $H$ such that $\exp\left(
X_{\cdot}^{1}\right)  $ and $\exp\left(  H_{\cdot}\right)  $ are martingales,
respectively. Since $X_{0}^{1}$ $=$ $H_{0}$ $=$ $0,$ we then have $\gamma$ $=$
$0$ in (\ref{gam1}).

Before considering affine processes with really unknown characteristic
function, in the next section we recall the known characteristics of a
log-Heston type model.

\subsection{The Heston model}

Let us consider for $X$ a log-Heston type model with dynamics%
\begin{align}
dX^{1}  &  =-\frac{1}{2}\alpha^{2}X^{2}dt+\alpha\sqrt{X^{2}}dW,\text{
\ \ \ }X^{1}(0)=0,\label{Hes}\\
dX^{2}  &  =\kappa\left(  \theta-X^{2}\right)  dt+\sigma\sqrt{X^{2}}\left(
\rho dW+\sqrt{1-\rho^{2}}d\overline{W}\right)  ,\text{\ \ \ }X^{2}%
(0)=\theta,\nonumber
\end{align}
for some $\alpha,\sigma,\kappa,\theta>0,$ and $-1\leq\rho\leq1.$ Note that the
initial value of $X^{2}$ is taken to be the expectation of the long-run
stationary distribution of $X^{2}.$ The characteristic function $X^{1}$ due to
(\ref{Hes}) is known as follows (we take Lord and Kahl's representation
\cite{LK}, due to the principal branch of the square root and
logarithm\footnote{Roger Lord confirmed to J.S. a typo in the published
version and so we refer to the preprint version.}):%
\begin{equation}
\ln\widehat{p}(t,\theta,z):=\ln\widehat{p}(t,\left(  0,\theta\right)  ,\left(
z,0\right)  )=A(z;t)+B(z;t)\theta,\text{ \ \ with} \label{logH}%
\end{equation}%
\begin{align}
A(z;t)  &  :=\frac{\theta\kappa}{\sigma^{2}}\left(  (a-d)t-2\ln\frac
{e^{-dt}-g}{1-g}\right)  ,\nonumber\\
B(z;t)  &  :=\frac{a+d}{\sigma^{2}}\frac{1-e^{dt}}{1-ge^{dt}}\text{
\ \ \ with}\label{Ric}\\
a  &  :=\kappa-\mathfrak{i}z\alpha\sigma\rho,\text{ \ \ }d:=\sqrt{a^{2}%
+\alpha^{2}\sigma^{2}\left(  \mathfrak{i}z+z^{2}\right)  },\text{
\ \ }g:=\frac{a+d}{a-d},\nonumber
\end{align}
while abusing notation in (\ref{logH}) slightly. By construction, $\exp\left(
X_{t}^{1}\right)  $ is a martingale and so it holds that $\ln\widehat
{p}(t,\theta,-\mathfrak{i})=0.$ This can be easily seen from the Heston
dynamics (\ref{Hes}) and also by taking $z=-\mathfrak{i}$ in (\ref{Ric}),
where we then have that $a=\kappa-z\alpha\sigma\rho\in\mathbb{R},$ so
$d=\left\vert a\right\vert .$ Thus $\left\vert g\right\vert =\infty$ if $a>0$
and $\left\vert g\right\vert =0$ if $a<0$ and for both cases we get that
$A(-\mathfrak{i};t)$ $\equiv$ $B(-\mathfrak{i};t)$ $\equiv$ $0.$ As a
consequence we have $\gamma$ $=$ $-\sigma^{2}/2$ in (\ref{gam}).

The generator (\ref{Gen}) due to the Heston model (\ref{Hes}) and its
corresponding symbol derivatives (\ref{sym}), i.e. the ingredients of the
recursion (\ref{Rec}), are spelled out in Appendix~\ref{AppA}.

\subsection{Heston model with state dependent jumps}

We now consider a generalized Heston model with state dependent jumps in the
first component, henceforth termed the HSDJ model, of the following form:%
\begin{align}
dX^{1}  &  =-\lambda_{0}a_{0}dt-\left(  \lambda_{1}a_{1}+\frac{1}{2}\alpha
^{2}\right)  X^{2}dt+\alpha\sqrt{X^{2}}dW\label{HSDJ}\\
&  +\int_{\mathbb{R}}y\left(  N(X_{-}^{2},dt,dy)-\lambda_{0}\mu_{0}%
(y)dydt-X^{2}\lambda_{1}\mu_{1}(y)dydt\right)  ,\nonumber\\
dX^{2}  &  =\kappa\left(  \theta-X^{2}\right)  dt+\sigma\sqrt{X^{2}}\left(
\rho dW+\sqrt{1-\rho^{2}}d\overline{W}\right) \nonumber
\end{align}
with $X^{1}(0)=0,$ $X^{2}(0)=\theta$ and with $t$ suppressed in $X_{t-}$ (cf.
(\ref{ItoLev})). In this model $N(w,dt,dy)$ is for each $w>0$ a Poisson point
process on $\mathbb{R}_{+}\times\mathbb{R}$ and $\mu_{0}$ and $\mu_{1}$ are
considered to be probability densities of jumps that arrive at rate
$\lambda_{0}>0$ and $w\lambda_{1}>0,$ respectively. Further in (\ref{HSDJ}),
$a_{0}$ and $a_{1}$ are non-negative constants given by%
\begin{equation}
a_{0}=\int\left(  e^{y}-y-1\right)  \mu_{0}(y)dy\text{ \ \ and \ \ }a_{1}%
=\int\left(  e^{y}-y-1\right)  \mu_{1}(y)dy, \label{ad}%
\end{equation}
hence in particular it is assumed that the measures associated with $\mu_{0}$
and $\mu_{1}$ have exponential moments. In the HSDJ model the density $\mu
_{0}$ may have support $\mathbb{R},$ for example Gaussian, while the density
$\mu_{1}$ may be concentrated on $(-\infty,0)$ for example. In this way
$\lambda_{0}$ and $\mu_{0}$ are responsible for the \textquotedblleft
normal\textquotedblright\ random jumps in (\ref{HSDJ}), while $\lambda_{1}$
and $\mu_{1}$ are responsible for downward jumps which, due to the (state)
dependence on $X^{2},$ arrive with increasing intensity as the volatility
$X^{2}$ increases. As such the model covers a stylized empirical fact observed
for several underlying quantities, such as assets, indices, or interest rates.
Since $\mu_{0}$ and $\mu_{1}$ are assumed to be probability densities, the
dynamics of $X^{1}$ in (\ref{HSDJ}) may also be written as%
\begin{align}
dX^{1}  &  =\left(  -\lambda_{0}\left(  \mathfrak{m}_{0}+a_{0}\right)
-\left(  \frac{1}{2}\alpha^{2}+\lambda_{1}\left(  \mathfrak{m}_{1}%
+a_{1}\right)  \right)  X^{2}\right)  dt\nonumber\\
&  +\alpha\sqrt{X^{2}}dW+\int_{\mathbb{R}}yN(X_{-}^{2},dt,dy), \label{dy2}%
\end{align}
with%
\begin{equation}
\mathfrak{m}_{0}:=\int_{\mathbb{R}}y\mu_{0}(y)dy\text{ \ \ and \ \ }%
\mathfrak{m}_{1}:=\int_{\mathbb{R}}y\mu_{1}(y)dy. \label{mom}%
\end{equation}
One can show rigorously that $e^{X_{t}^{1}}$ is a martingale with
$\mathbb{E}\left[  e^{X_{t}^{1}}\right]  =1,$ and so we may take in
(\ref{gam}) $\gamma=-\sigma^{2}/2$ again. Here we restrict our selves to a
heuristic argumentation: From It\^{o}'s formula for jump processes we see that
for $0\leq u<t,$%
\begin{align}
e^{X_{t}^{1}}-e^{X_{u}^{1}}  &  =\int_{u}^{t}e^{X_{s-}^{1}}d\left(  X_{s}%
^{1}\right)  ^{\text{cont.}}+\frac{1}{2}\int_{u}^{t}e^{X_{s-}^{1}}\alpha
^{2}X_{s}^{2}ds+\sum_{u<s\leq t}\left\{  e^{X_{s}^{1}}-e^{X_{s-}^{1}}\right\}
\label{It}\\
&  =\int_{u+}^{t}e^{X_{s-}^{1}}\left(  \left(  -\lambda_{0}\left(
\mathfrak{m}_{0}+a_{0}\right)  -\lambda_{1}\left(  \mathfrak{m}_{1}%
+a_{1}\right)  X^{2}\right)  dt+\alpha\sqrt{X_{s}^{2}}dW\right) \nonumber\\
&  +\sum_{u<s\leq t}\left\{  e^{X_{s}^{1}}-e^{X_{s-}^{1}}\right\}  .\nonumber
\end{align}
So, heuristically, we have that%
\begin{gather*}
\mathbb{E}\left[  \left.  \sum_{u<s\leq u+\Delta}\left\{  e^{X_{s}^{1}%
}-e^{X_{s-}^{1}}\right\}  \right\vert X_{u-}\right]  \approx e^{X_{u-}^{1}%
}\mathbb{E}\left[  \left.  \sum_{u<s\leq u+\Delta}\left\{  e^{X_{s}^{1}%
-X_{s-}^{1}}-1\right\}  \right\vert X_{u-}\right] \\
=e^{X_{u-}^{1}}\,\lambda_{0}\Delta\int\left(  e^{y}-1\right)  \mu
_{0}(y)dy+X_{u-}^{2}e^{X_{u-}^{1}}\lambda_{1}\Delta\int\left(  e^{y}-1\right)
\mu_{1}(y)dy\\
=e^{X_{u-}^{1}}\,\left(  \lambda_{0}\left(  \mathfrak{m}_{0}+a_{0}\right)
+X_{u-}^{2}\lambda_{1}\left(  \mathfrak{m}_{1}+a_{1}\right)  \right)  \Delta,
\end{gather*}
for $\Delta\downarrow0.$ Combining with (\ref{It}) this yields
\[
e^{X_{u+\Delta}^{1}}-e^{X_{u}^{1}}\approx e^{X_{u-}^{1}}\alpha\sqrt{X_{u}^{2}%
}\Delta W+\zeta_{u,u+\Delta}%
\]
with $\mathbb{E}\left[  \left.  \zeta_{u,u+\Delta}\right\vert X_{u-}\right]
=0.$

In Appendix~\ref{AppB} we spell out the generator, cf. (\ref{Gen}), and its
corresponding symbol derivatives (\ref{sym}) corresponding to the HSDJ model
(\ref{HSDJ}).

\begin{example}
In the case where $\lambda_{1}=0,$ the characteristic function $\widehat
{p}_{\lambda_{0},\mu_{0}}$ of $X^{1}$ is simply given by (see (\ref{dy2}),
(\ref{ad}) and (\ref{mom}))%
\begin{align*}
\ln\widehat{p}_{\lambda_{0},\mu_{0}}(t,\theta,z)  &  =\ln\widehat{p}%
(t,\theta,z)-t\lambda_{0}\left(  a_{0}+\mathfrak{m}_{0}\right)  \mathfrak{i}%
z+t\lambda_{0}\psi_{0}(z)\\
&  =\ln\widehat{p}(t,\theta,z)-t\lambda_{0}\psi_{0}(-\mathfrak{i}%
)\mathfrak{i}z+t\lambda_{0}\psi_{0}(z),
\end{align*}
where%
\[
\psi_{0}(z):=\int\left(  e^{\mathfrak{i}yz}-1\right)  \mu_{0}(y)dy=\int
e^{\mathfrak{i}yz}\mu_{0}(y)dy-1
\]
follows from the characteristic function of the jump measure and $\widehat
{p}(t,\theta,z)$ is given by (\ref{logH}). Note that we have $\ln\widehat
{p}_{\lambda_{0},\mu_{0}}(t,\theta,-\mathfrak{i})=0$ again indeed. For example
if the jumps are $\mathcal{N}(c,\nu^{2})$ distributed we have the well known
expression%
\[
\psi_{0}(z)=e^{\mathfrak{i}cz-\frac{1}{2}\nu^{2}z^{2}}-1,
\]
hence%
\[
\ln\widehat{p}_{\lambda_{0},c,\nu^{2}}(t,\theta,z):=\ln\widehat{p}%
(t,\theta,z)+t\lambda_{0}\left(  e^{\mathfrak{i}cz-\frac{1}{2}\nu^{2}z^{2}%
}-\mathfrak{i}ze^{c+\frac{1}{2}\nu^{2}}+\mathfrak{i}z-1\right)  .
\]

\end{example}

\section{Numerical examples}

In this section we will price European options by a Fourier based method due
to Carr-Madan \cite{CM}. Let the stock price at maturity $T$ be given as%
\[
S_{T}=S_{0}e^{rT+Y_{T}},
\]
where $\exp\left[  Y_{\cdot}\right]  $ is a martingale with $Y_{0}=0.$ If the
characteristic function
\[
\Phi_{T}(z):=\mathbb{E}\,\left[  e^{\mathfrak{i}zY_{T}}\right]
\]
is known, then the the price of a European call option with strike $K$ at time
$t$ $=$ $0$ is given by
\begin{equation}
C(K)=(S_{0}-Ke^{-rT})^{+}+\frac{S_{0}}{2\pi}\int_{-\infty}^{\infty}%
\frac{1-\Phi_{T}(z-\mathfrak{i})}{z(z-\mathfrak{i})}e^{-\mathfrak{i}z\ln
\frac{Ke^{-rT}}{S_{0}}}dz \label{CM}%
\end{equation}
(Carr-Madan's formula). For more general Fourier valuation formulas, see
\cite{EGP}. In general, the decay of the integrand in (\ref{CM}) is of order
$O(\left\vert z\right\vert ^{-2})$ as $\left\vert z\right\vert \rightarrow
\infty,$ hence relatively slow. We therefore use a kind of variance reduction
for integrals using the formula%
\begin{equation}
\mathcal{BS}\left(  S_{0},T,r,\sigma_{B}\right)  =(S_{0}-Ke^{-rT})^{+}%
+\frac{S_{0}}{2\pi}\int_{-\infty}^{\infty}\frac{1-\Phi_{T}^{\mathcal{BS}%
}(z-\mathfrak{i})}{z(z-\mathfrak{i})}e^{-\mathfrak{i}z\ln\frac{Ke^{-rT}}%
{S_{0}}}dz, \label{CM1}%
\end{equation}
where $\mathcal{BS}$ is the well-known Black-Scholes formula based on the
risk-neutral Black-Scholes model
\begin{align*}
S_{t}^{\mathcal{B}}  &  :=S_{0}e^{rT-\sigma_{B}^{2}T/2+\sigma_{B}W_{T}},\text{
\ \ with}\\
\Phi_{T}^{\mathcal{BS}}(z)  &  :=\mathbb{E}\,\left[  e^{\mathfrak{i}z\left(
-\sigma_{B}^{2}T/2+\sigma_{B}W_{T}\right)  }\right]  =e^{-\left(
z^{2}+\mathfrak{i}z\right)  \sigma_{B}^{2}T/2},
\end{align*}
for a suitable but in principle arbitrary $\sigma_{B}>0.$ Next, subtracting
(\ref{CM}) and (\ref{CM1}) gives the variance reduced formula%
\begin{equation}
C(K)=\mathcal{BS}\left(  S_{0},T,r,\sigma_{B}\right)  +\frac{S_{0}}{2\pi}%
\int_{-\infty}^{\infty}\frac{\Phi_{T}^{\mathcal{BS}}(z-\mathfrak{i})-\Phi
_{T}(z-\mathfrak{i})}{z(z-\mathfrak{i})}e^{-\mathfrak{i}z\ln\frac{Ke^{-rT}%
}{S_{0}}}dz,\label{eq:var-red-fourier-pricing}
\end{equation}
where, typically, the integrand decays much faster than in (\ref{CM}).

\subsection{Product of Heston models}
\label{sec:prod-hest-models}

We first consider a model where the stock price $S_t$ is obtained as the
product of two independent Heston factors, i.e.,~\eqref{glnx1} with $X^1_t$
another Heston model. Clearly, in this case a closed form expression for the
characteristic function of $\ln S_t$ exists, and therefore the asymptotic
expansion presented in this paper is not needed for pricing. This allows us to
easily compute accurate reference prices, and thus assess the numerical
accuracy of prices obtained from the expansion of the characteristic
function. All calculations were done using Mathematica. Using its symbolic
capabilities, we have implemented the recursion~\eqref{Rec} in full
generality.

\begin{table}[!htp]
  \centering
  \begin{tabular}{c|cc}
    & $H_t$ & $X^1_t$\\
    \hline
    $\alpha$ & $1.0$ & $1.0$ \\
    $\kappa$ & $1.5$ & $1.5$ \\
    $\sigma$ & $0.6$ & $0.3$ \\
    $\theta$ & $0.04$ & $0.0225$ \\
    $\rho$ & $-0.2$ & $-0.3$ \\
    $v$ & $0.04$ & $0.0225$ \\
  \end{tabular}
  \caption{Parameters of the Heston+Heston-model. $v$ denotes the initial
    variance in both components.}
  \label{tab:parameters-HH}
\end{table}

The Heston parameters for the components $H_t$ and $X^1_t$ are presented in
Table\ref{tab:parameters-HH}. Additionally, we choose $S_0 = 10$ and $r =
0.05$ for option pricing. Based on these parameters, we compute the asymptotic
expansion $\widehat{p}_K$ of the characteristic function using~\eqref{boden}
with $K = 8$, i.e., including the first \emph{nine} terms in the expansion.

\begin{figure}[!htp]
  \centering
  \begin{subfigure}[b]{0.5\textwidth}
    \centering
    \includegraphics[width=\textwidth]{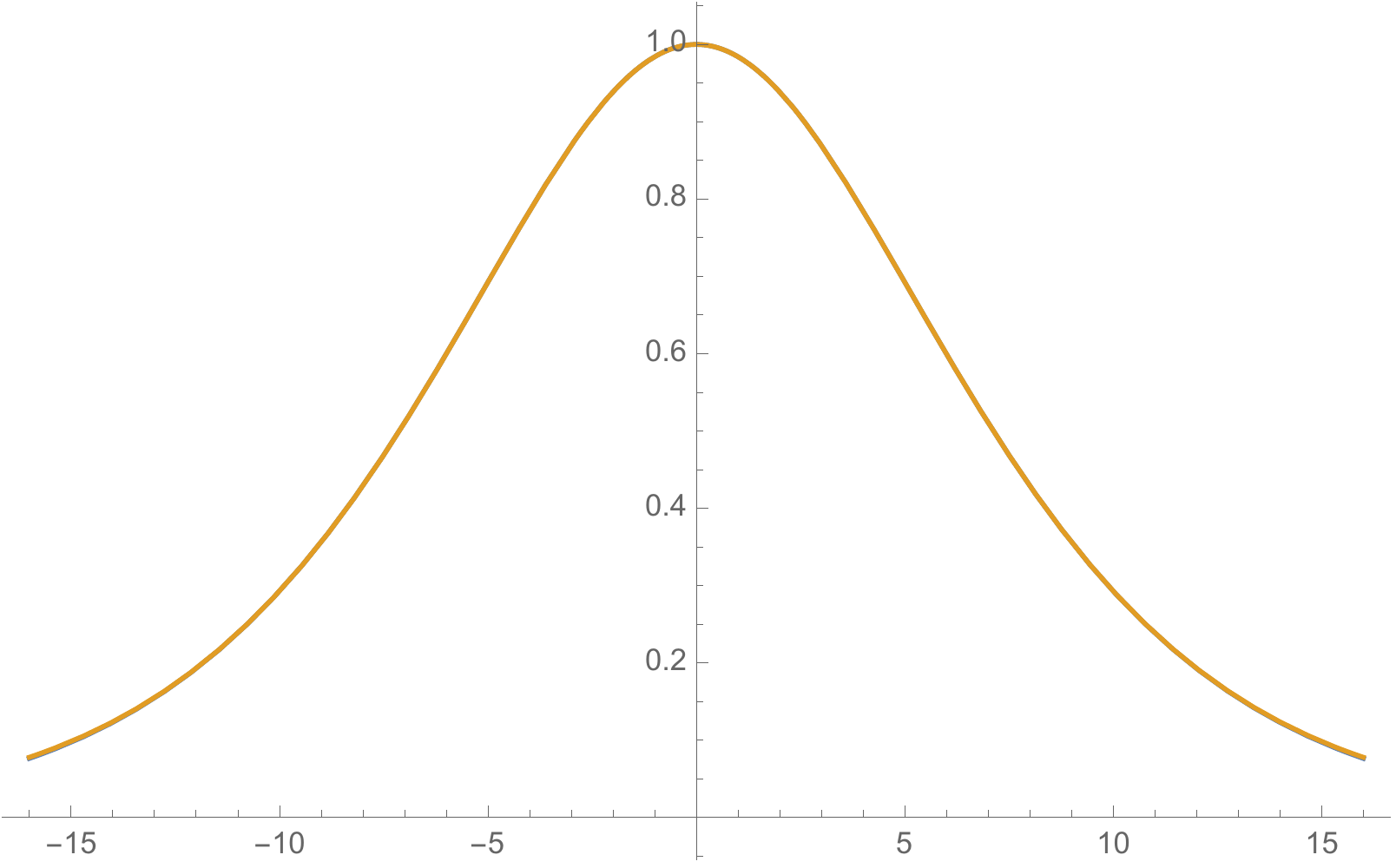}
    \caption{Real part, $t=1/2$}
  \end{subfigure}%
  ~
  \begin{subfigure}[b]{0.5\textwidth}
    \centering
    \includegraphics[width=\textwidth]{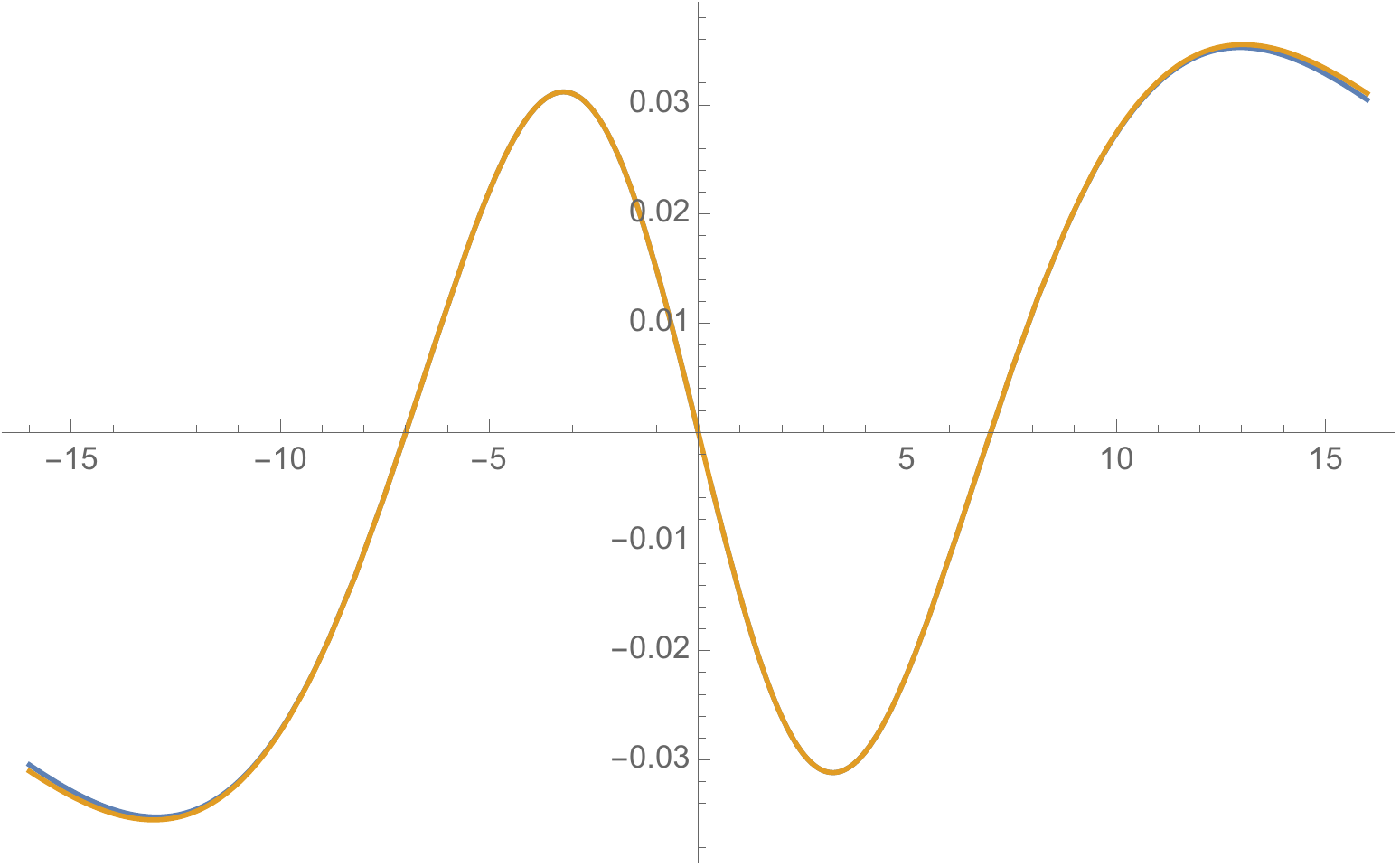}
    \caption{Imaginary part, $t=1/2$}
  \end{subfigure}

  \begin{subfigure}[b]{0.5\textwidth}
    \centering
    \includegraphics[width=\textwidth]{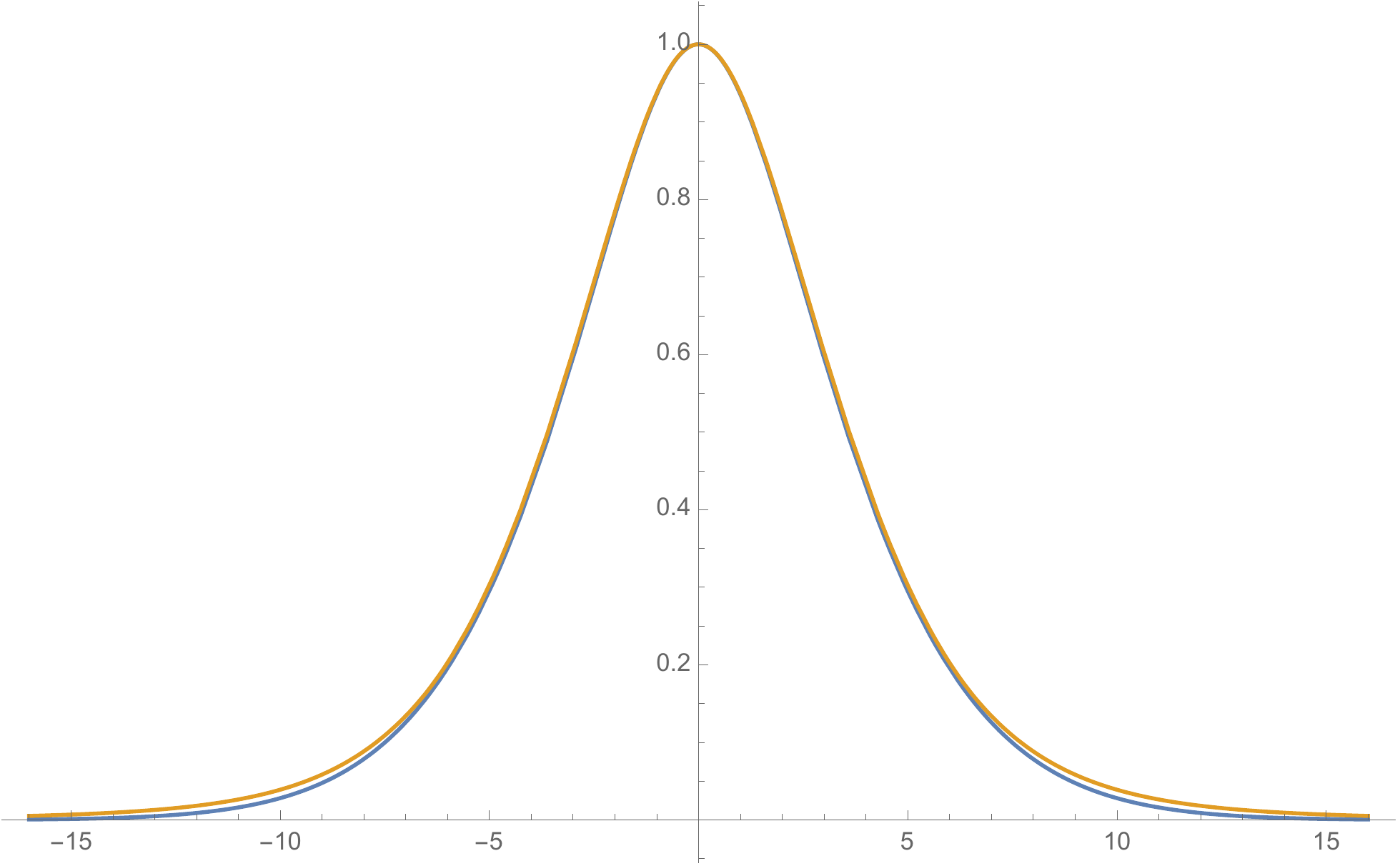}
    \caption{Real part, $t=2$}
  \end{subfigure}%
  ~
  \begin{subfigure}[b]{0.5\textwidth}
    \centering
    \includegraphics[width=\textwidth]{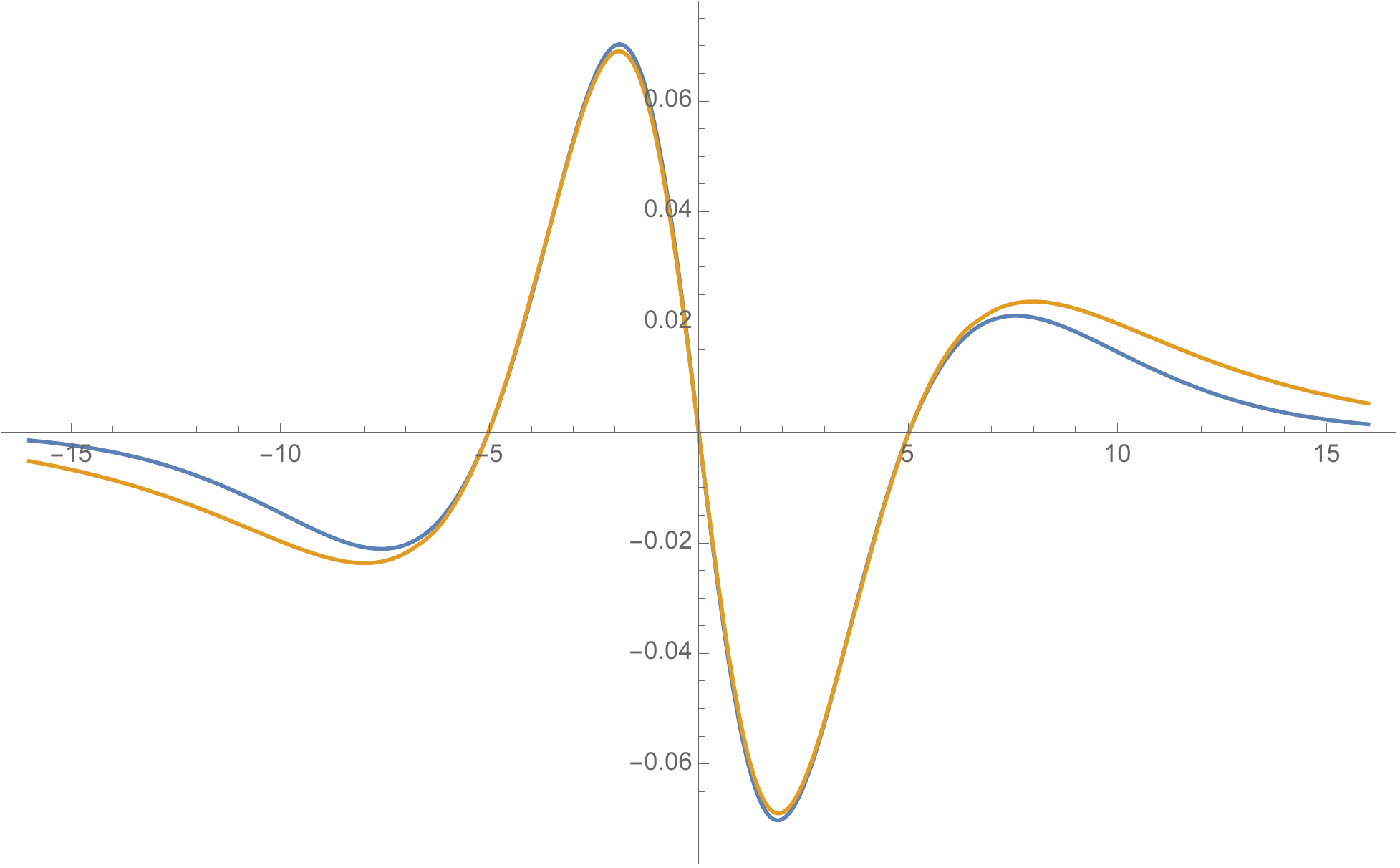}
    \caption{Imaginary part, $t=2$}
  \end{subfigure}
  \caption{Exact (blue) and approximate (orange) characteristic functions of
    the logarithm of the normalized stock price in the generalized Merton
    model with two Heston factors evaluated at time $t=1/2$ and $t=2$
    (years).}
  \label{fig:char-HH}
\end{figure}

In Figure~\ref{fig:char-HH}, we compare the exact and the approximate
characteristic functions of the (normalized) logarithm of the stock
prices---i.e., with $S_0 = 1$ for convenience. We can clearly see that the
approximation deteriorates when $|u|$ becomes large, but then both the
exact and the approximate characteristic functions tend to $0$. Moreover, the
approximation formula is more accurate for small $t$.

\begin{table}[!htp]
  \centering
  \begin{tabular}{l|cccccc}
    $L$ & $2$ & $4$ & $8$ & $16$ & $32$ & $64$ \\
    \hline
    Exact & $0.8350$ & $0.9621$ & $1.1105$ & $1.1832$ &  $1.1884$ &
    $1.884$ \\
    Approx. & $0.8353$ & $0.9626$ & $1.1111$ & $1.1842$ & $1.1896$ &
    $1.1896$\\
    (Rel.~error) & $0.2981$ & $0.1912$ & $0.0665$ & $0.0054$ &
    $0.0010$ & $0.0010$\\
  \end{tabular}
  \caption{Price of ATM call option with maturity $T = 1$ computed using
    domain of integration $[-L,L]$ for both the exact characteristic function
    and the approximate formula, together with the relative error for using
    the approximate formula---w.r.t.~the most accurate price obtained from the
  exact formula.}
  \label{tab:HH-int-dom}
\end{table}

When we come to option pricing, we plug the approximate formula for the
characteristic function into the Fourier pricing
formula~\eqref{eq:var-red-fourier-pricing}. For the implementation, we clearly
need to replace the infinite domain of integration by a finite one, i.e., we
use~\eqref{eq:var-red-fourier-pricing} integrating from $-L$ to $L$, $L \in
\mathbb{R}$. This cut-off is potentially critical for our approximation
procedure, as large integration domains (and, hence, large $|u|$) may
correspond to large errors of the approximate formula. Fortunately,
Table~\ref{tab:HH-int-dom} indicates that this effect does not materialize.

\begin{remark}
  At this stage, we would like to highlight once more the heuristic choice of
  $\eta$ proposed in Remark~\ref{rem:heuristic-eta}. Without a good choice of
  $\eta$, it is very easy to run into situations, where the approximation
  error is already too large for the needed domain of integration.
\end{remark}

\begin{figure}[!htp]
  \centering
  \begin{subfigure}[b]{0.5\textwidth}
    \centering
    \includegraphics[width=\textwidth]{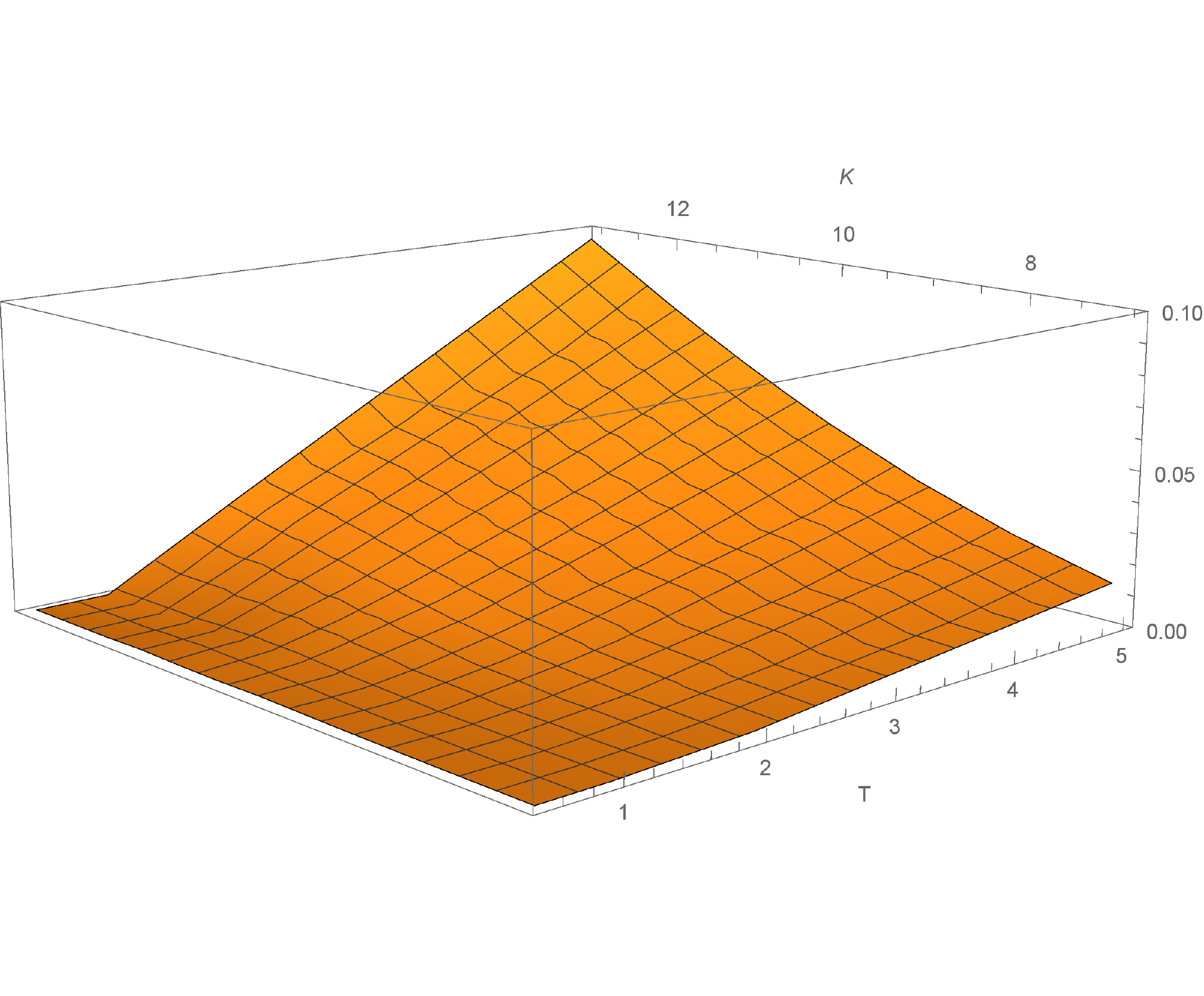}
    \caption{Relative error.}
  \end{subfigure}%
  ~
  \begin{subfigure}[b]{0.5\textwidth}
    \centering
    \includegraphics[width=\textwidth]{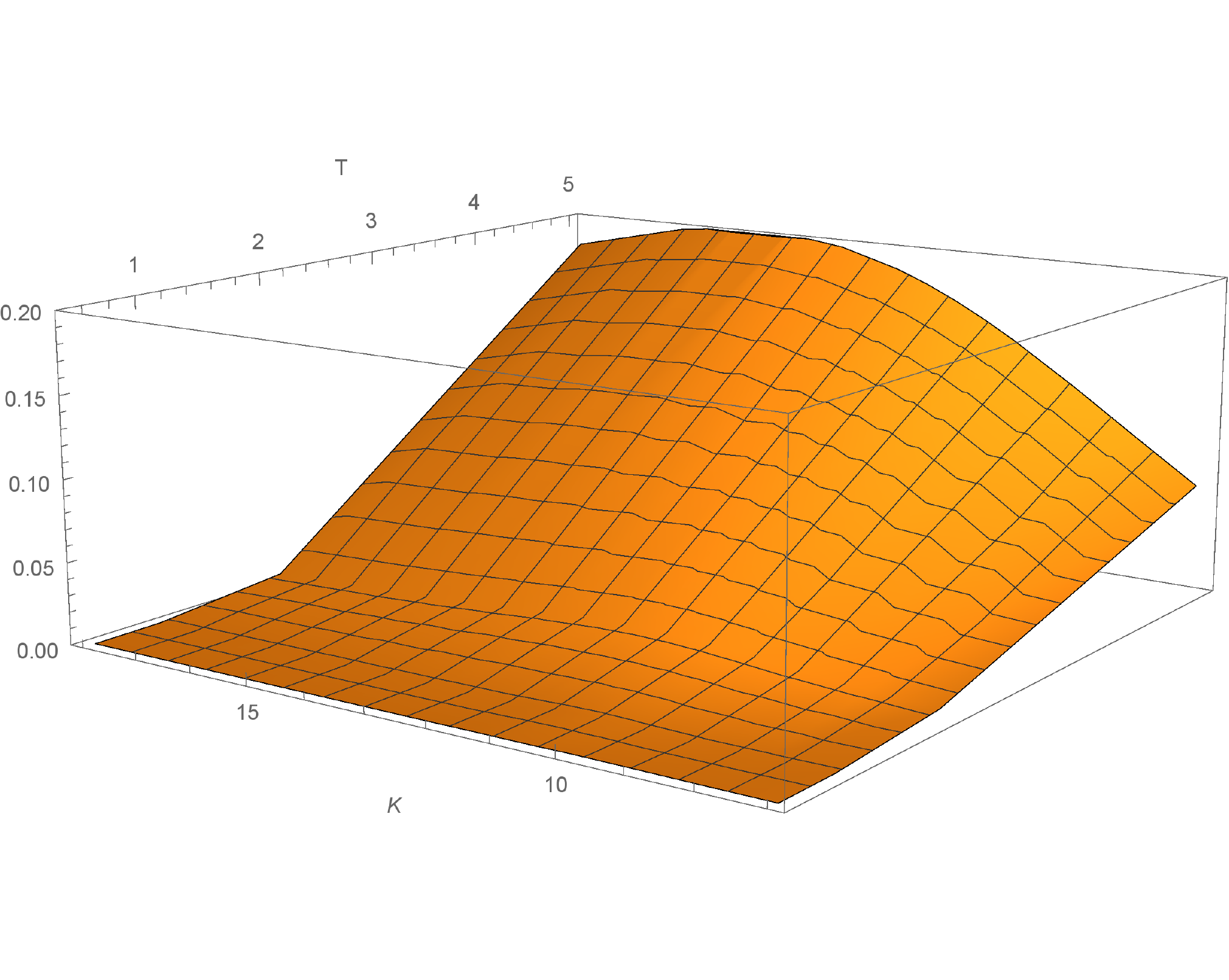}
    \caption{Absolute error.}
  \end{subfigure}
  \caption{Relative and absolute errors of European call option prices.}
  \label{fig:HH-error}
\end{figure}

Let us consider option prices and the corresponding errors for
maturities from $1/2$ to $5$ years and for strike prices between $7$ (deep in)
and $13$ (deep out of) the money. Figure~\ref{fig:HH-error} shows that errors
remain small ($\le 2 \%$ ATM) for maturities up to $2$ years. For (deep) OTM
options, it seems to be more reasonable to look at absolute instead of
relative errors, which give a similar impression.

\begin{figure}[!htp]
  \centering
  \includegraphics[width=0.5\textwidth]{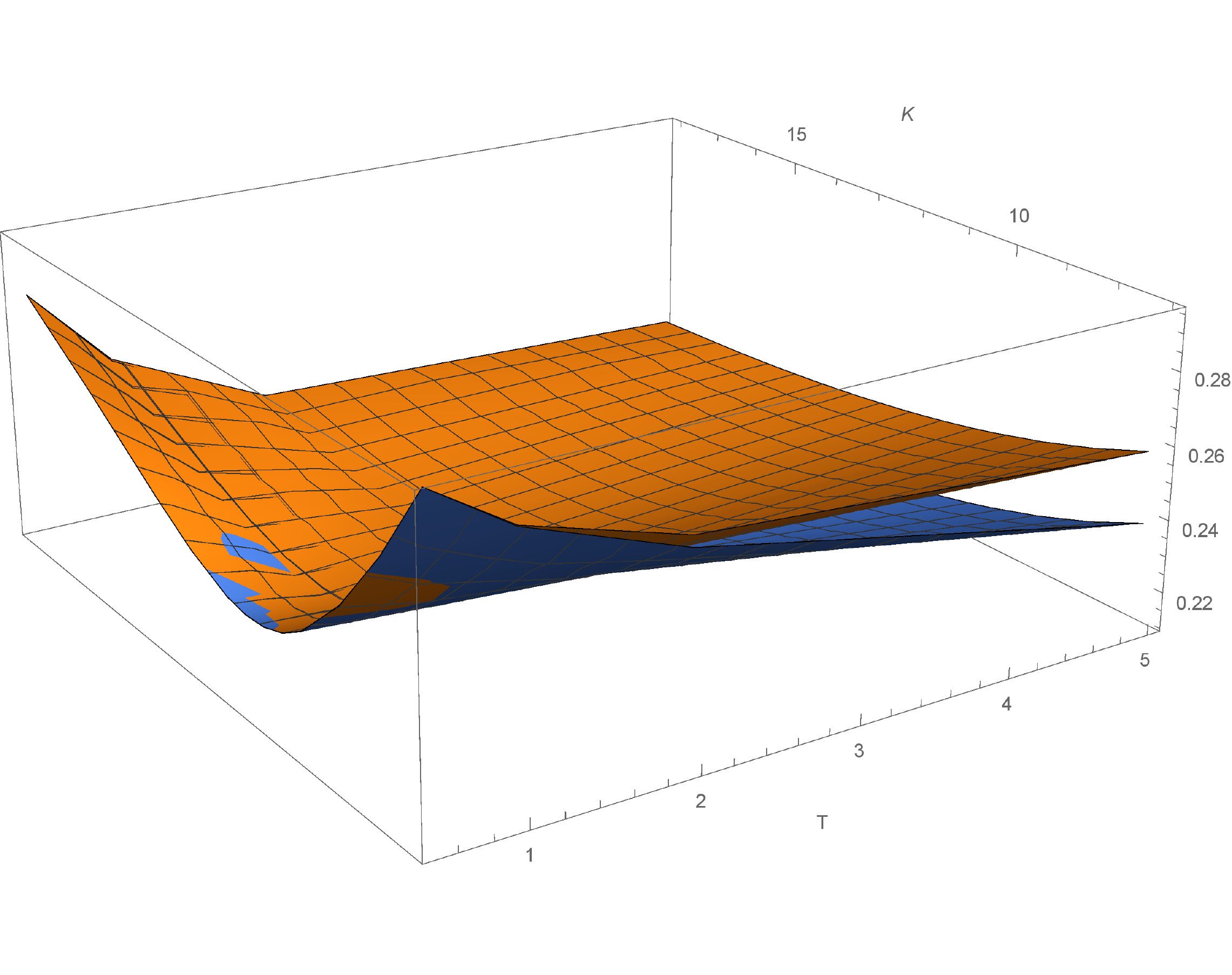}
  \caption{Implied volatility of the generalized Merton model based on two
    Heston factors based on exact (blue) and approximate (orange)
    characteristic functions.}
  \label{fig:HH-iv}
\end{figure}

Finally, the implied volatility in this model is plotted in
Figure~\ref{fig:HH-iv}. Considerable deviations between the exact and the
approximate formula are only observed for higher maturities.

\subsection{Generalized Merton model with state-dependent jumps}
\label{sec:gener-mert-model}

Let us consider a generalized Merton model of the form (\ref{glnx1}) where
$X^1$ is an affine jump process with state-dependent jump-intensity In the
sense of~\eqref{HSDJ}. The parameters corresponding to the diffusive parts of
both $H$ and $X^1$ are chosen as in Table~\ref{tab:parameters-HH}. Regarding
the jump part of $X^1$, we set $\lambda_0 = 0$, $\mu_0 = 0$, thereby turning
off the jumps with constant, i.e., not state dependent, intensities. The jump
parameters of $X^1$ are chosen according to Table\ref{tab:HHJ-par}.

% \begin{table}[!htbp]
%   \centering
%   \begin{tabular}{c|cc}
%     & $H_t$ & $X^1_t$\\
%     \hline
%     $\alpha$ & $1.0$ & $1.0$\\
%     $\kappa$ & $1.5$ & $1.5$
%   \end{tabular}
%   \caption{Diffusion parameters of Heston model with jumps in the second factor. $v$ denotes the initial variance terms.}
%   \label{tab:parameters-HHJ-diff}
% \end{table}

\begin{table}[!htp]
  \centering
  \begin{tabular}{c|c}
    & $X^1_t$ \\
    \hline
    $\lambda_1$ & $10$\\
    $\mu_1(y)$ & $\mathbf{1}_{y<0} p e^{py}$\\
    $p$ & $4.48$
  \end{tabular}
  \caption{Jump parameters of $X^1$}
  \label{tab:HHJ-par}
\end{table}

\begin{figure}[!htbp]
  \centering
  \includegraphics[width=0.7\textwidth]{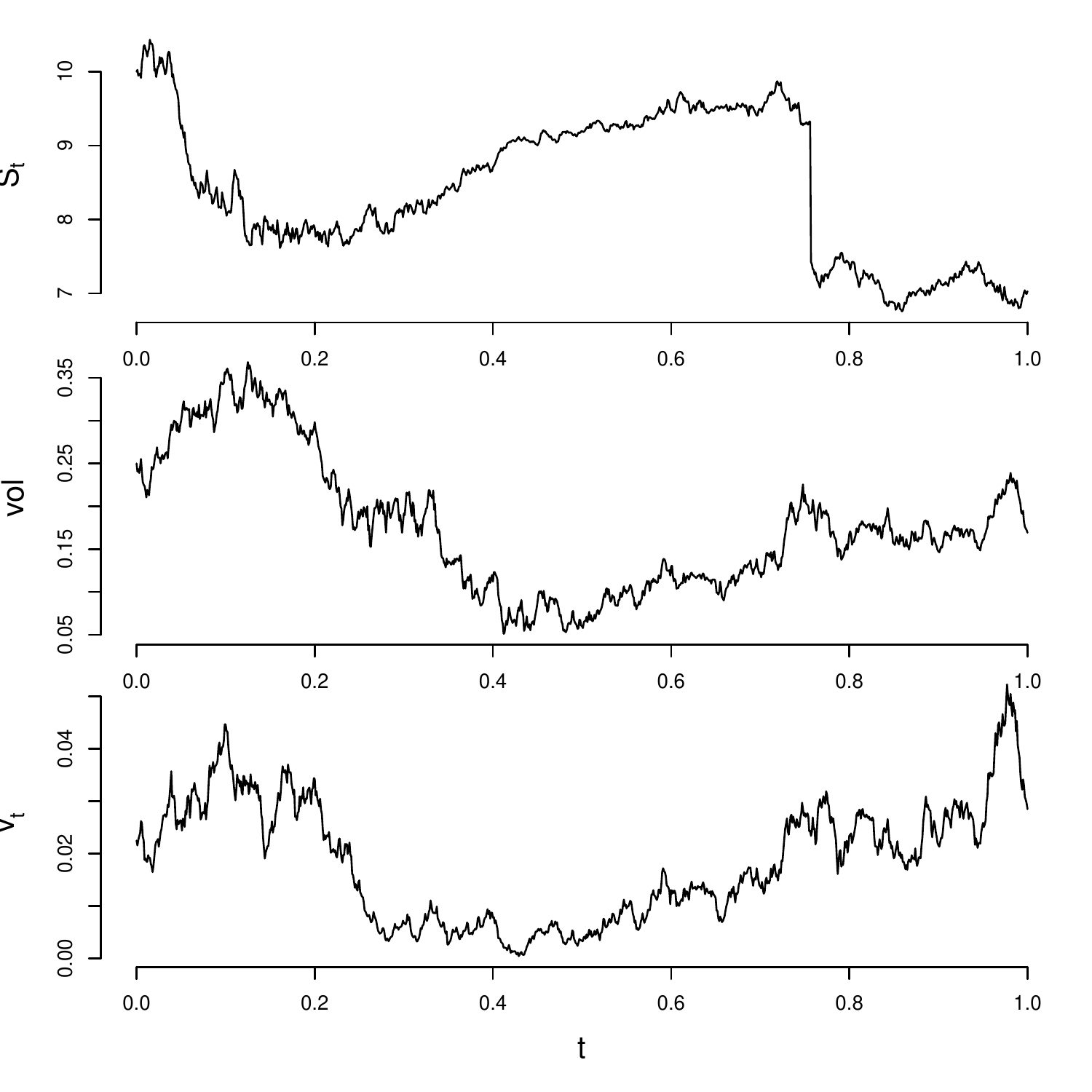}
  \caption{Sample path of $S_t$ in the generalized Merton model with
    state-dependent jumps (first panel), volatility (more precisely,
      the square root of the sum of both variance components) of $S_t$
    (second panel), and of the variance component of the second Heston
    factor. A jump occurs shortly after time $0.75$.}
  \label{fig:HJ-sample}
\end{figure}

This means that jumps in the log-price have exponentially distributed
magnitude and negative sign. The mean jump of the log-price is around $0.22$,
i.e., in case of a downward jump (``crisis''), the stock loses about $20\%$ of
its value. The intensity $\lambda_1$ seems excessively high, but recall that
this intensity is multiplied by the instantaneous variance of the Heston
component, which is started at $0.04$.

By (\ref{defp}), and (\ref{defp1}) below, we obtain
\begin{align*}
\psi_{0}(\xi) &  =0,\text{ \ \ \ \ \ }\mathfrak{m}_{0}+a_{0}=\psi
_{0}(-\mathfrak{i})=0,\\
\psi_{1}(\xi) &  =\int_{-\infty}^{0}\left(  e^{\mathfrak{i}\xi y}-1\right)
pe^{py}dy=-\frac{\mathfrak{i}\xi}{p+\mathfrak{i}\xi},\text{ \ \ \ \ \ }%
\mathfrak{m}_{1}+a_{1}=\psi_{1}(-\mathfrak{i})=-\frac{1}{p+1}.
\end{align*}

\begin{figure}[!htp]
  \centering
  \begin{subfigure}[b]{0.5\textwidth}
    \centering
    \includegraphics[width=\textwidth]{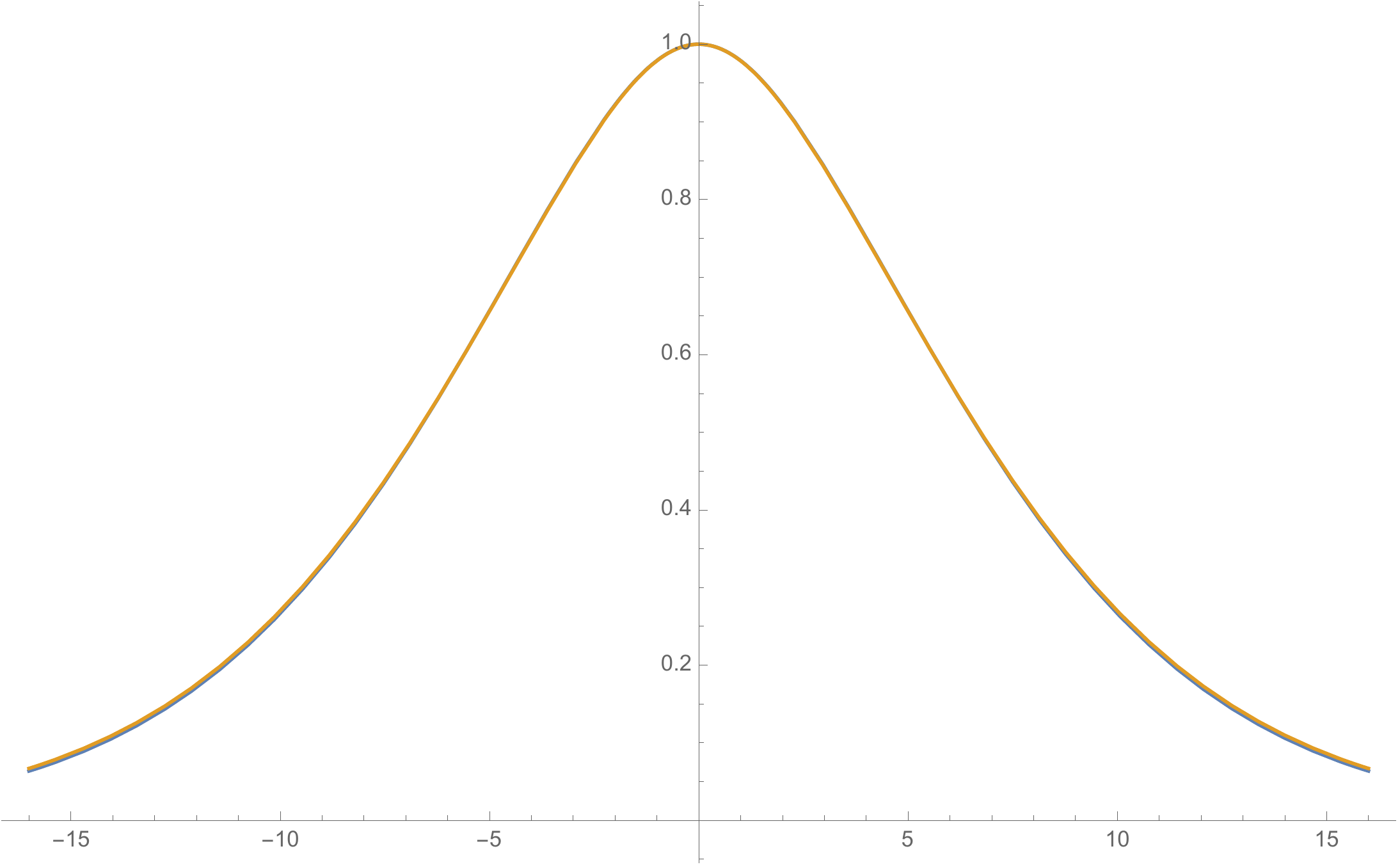}
    \caption{Real part, $t=1$}
  \end{subfigure}%
  ~
  \begin{subfigure}[b]{0.5\textwidth}
    \centering
    \includegraphics[width=\textwidth]{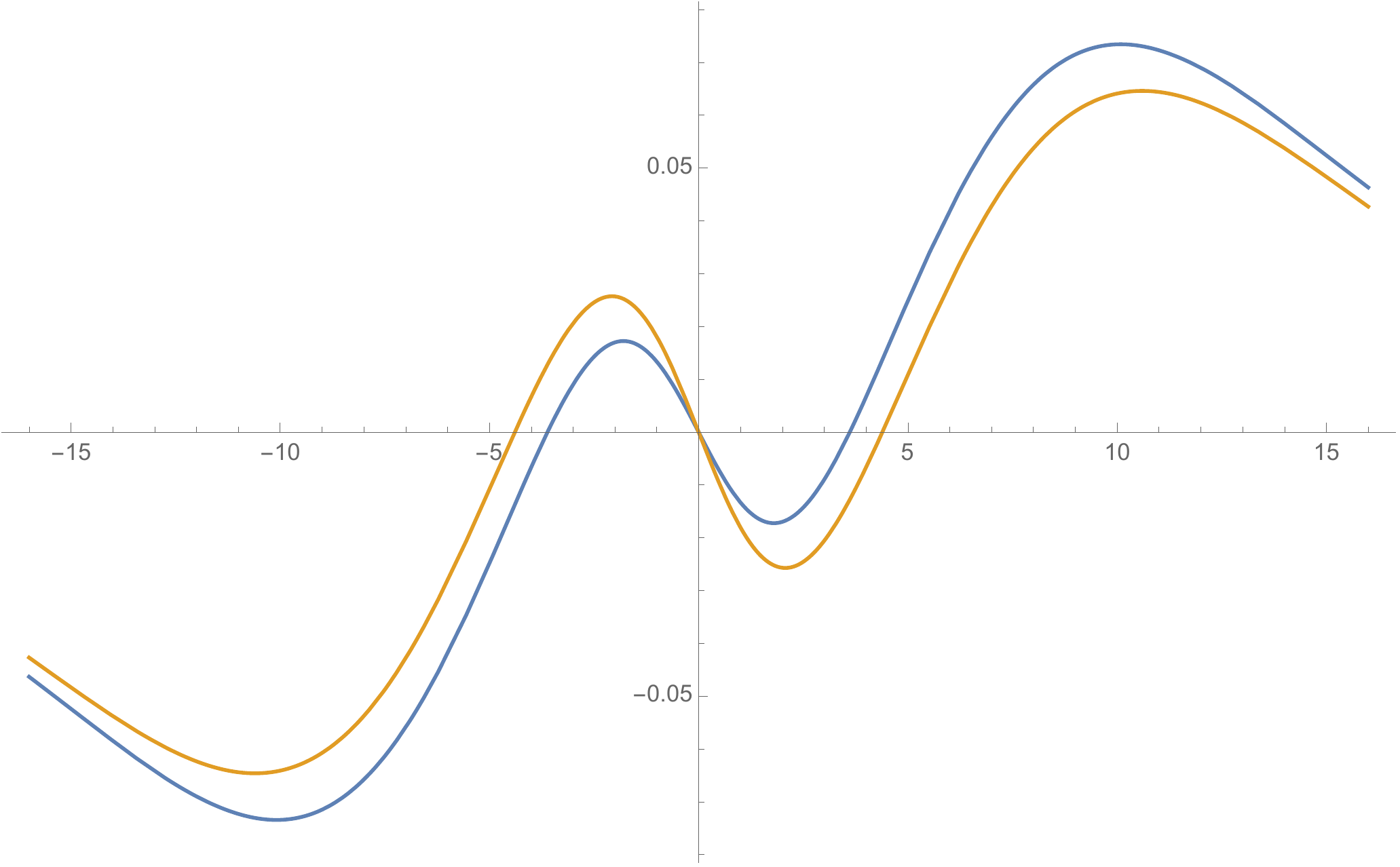}
    \caption{Imaginary part, $t=1$}
  \end{subfigure}
  \caption{Approximate characteristic function (orange) of the logarithm of
    the normalized stock price in the generalized Merton model with one Heston
    factor and one Heston factor with jumps evaluated at time $t=1/2$
    (year). Comparison with the characteristic function computed by a Monte
    Carlo simulation (blue).}
  \label{fig:char-HJ}
\end{figure}

Figure~\ref{fig:char-HJ} shows the approximate characteristic function
including jumps at time $t=1/2$, compared with the exact characteristic
function without jumps. As expected, the jumps lead to a considerable change
in the characteristic function. We compare the characteristic function to
another numerical approximation based on Monte Carlo simulation. Both
approximations lead to very close results especially in the real part. The
results are less close for the imaginary part, but notice that the graphical
representation exaggerates the differences as the scale is much smaller in the
second plot (from $-0.1$ to $0.1$ instead of $0$ to $1$).

\begin{figure}[!htb]
  \centering
  \includegraphics[width=0.5\textwidth]{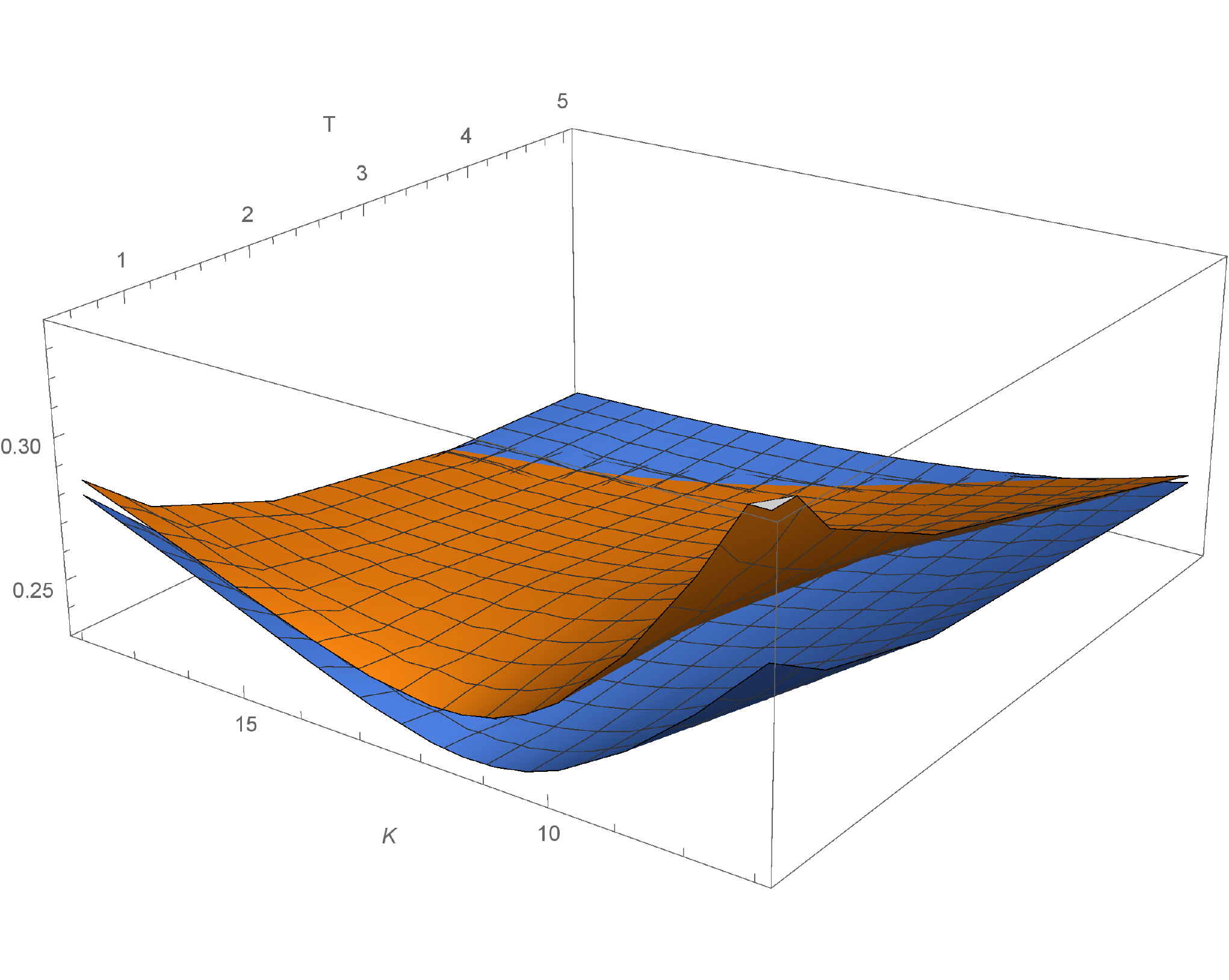}
  \caption{Implied volatility of the generalized Merton model with one Heston
    factor and one Heston factor with jumps (orange), compared with the
    implied volatilities computed with the exact characteristic function in
    Figure~\ref{fig:HH-iv}.}
  \label{fig:HJ-iv}
\end{figure}

These changes in the distribution have the expected changes in the option
prices. In particular, the implied volatilities become larger, and also the
smile becomes much more pronounced, comparing Figure~\ref{fig:HJ-iv} with
Figure~\ref{fig:HH-iv}.

\begin{table}[!htbp]
  \centering
  \begin{tabular}{l|ccccccc}
    $K$ & $7$ & $8$ & $9$ & $10$ & $11$ & $12$ & $13$\\
    \hline
    Monte Carlo & $3.2719$ & $2.3688$ & $1.5511$ &
    $0.8888$ & $0.4427$ & $0.2006$ & $0.0884$\\
    Asym.~formula & $3.2279$ & $2.3276$ & $1.5144$ & $0.8583$ &
    $0.4217$ & $0.1880$ & $0.0818$ \\
    Rel.~error & $0.0134$ & $0.0174$ & $0.0237$ & $0.0343$ & $0.0476$ &
    $0.0627$ & $0.0744$\\
    $\frac{\text{MC stat.~error}}{\text{Ref.~price}}$ & $0.0018$ & $0.0023$ &
    $0.0031$ & $0.0044$ & $0.0067$ & $0.0106$ & $0.0166$ \\
  \end{tabular}
  \caption{Option prices for maturity $T = 1/2$ for various strike prices in
    the Heston model plus jumps. We compare prices obtained by the asymptotic
    expansion of the characteristic function with prices obtained by Monte
    Carlo simulation.}
  \label{tab:prices-HHJ}
\end{table}

Finally, let us directly compare the price for some European call options with
reference prices obtained by Monte Carlo simulation, see
Table~\ref{tab:prices-HHJ}. Once again, we used $S_0 = 10$ and $r = 0.05$. The
Monte Carlo prices are based on $100,000$ trajectories with $1000$ time-steps
each, the statistical error, i.e., the standard deviation divided by the
square root of the number of samples, is considerable smaller than the
observed difference.

Unfortunately, the results of Table~\ref{tab:prices-HHJ} are not as
convincing as the accuracy of the approximation in the pure diffusion case
suggested, compare Table~\ref{tab:HH-int-dom} and
Figure~\ref{fig:HH-error}. We suspect a combination of slow decay of the
characteristic function, sub-optimal choice of the damping parameter $\eta$
and higher truncation error of the asymptotic characteristic function, see the
conclusions below for some further comments.

\paragraph{Conclusions}
\label{sec:conclusions}

From the examples we conclude that for times being not too large the
approximation procedure based on \cite{BKS09} performs rather well. More
specifically, if no jumps are in the play the procedure works very good, but
with incorporated (state dependent) jumps the accuracy is somewhat less. In
order to resolve this issue one could investigate different directions. One
reason for less accuracy may be a diminished effect of the Black-Scholes
ingredients in the Fourier pricing formula~\eqref{eq:var-red-fourier-pricing}
in the presence of state dependent jumps. This in turn might require a larger
integration range where that approximation gets worse at the upper and lower
end, respectively.  As a way out, it looks natural to replace the role of the
Black-Scholes ingredients in ~\eqref{eq:var-red-fourier-pricing} by an affine
model with state independent jumps for which the characteristic function is
known, leading to a representation of the form%
\begin{gather*}
C^{\text{appr}}(K)=(S_{0}-Ke^{-rT})^{+}+\frac{S_{0}}{2\pi}\int_{-\infty
}^{\infty}\frac{1-\Phi_{T}^{\text{known}}(z-\mathfrak{i})}{z(z-\mathfrak{i}%
)}e^{-\mathfrak{i}z\ln\frac{Ke^{-rT}}{S_{0}}}dz\\
+\frac{S_{0}}{2\pi}\int_{-\infty}^{\infty}\frac{\Phi_{T}^{\text{known}%
}(z-\mathfrak{i})-\Phi_{T}^{\text{appr}}(z-\mathfrak{i})}{z(z-\mathfrak{i}%
)}e^{-\mathfrak{i}z\ln\frac{Ke^{-rT}}{S_{0}}}dz=:I_{\text{known}%
}+I_{\text{appr}}.
\end{gather*}
The integral $I_{\text{known}}$ can be computed with any desired accuracy
while for the integral $I_{\text{appr}}$ a relatively small integration range
may be sufficient.

Other reasons for the decreased accuracy in Section~\ref{sec:gener-mert-model}
for instance, may be a too small $\eta$ chosen due to
Remark~\ref{rem:heuristic-eta}, or not enough iterations. However, we leave
all these investigations for further research, since this article is
considered merely a first guide on numerical implementation of the method in
\cite{BKS09}.

\appendix

\section{Generator and $\mathfrak{b}_{\beta}$ for the Heston model}

\label{AppA}

By conferring (\ref{Gen}), (\ref{ItoLev}), and (\ref{Hes}), the generator of
the Heston model is given by
\[
A=-\frac{1}{2}\alpha^{2}x_{2}\partial_{x_{1}}+\kappa\left(  \theta
-x_{2}\right)  \partial_{x_{2}}+\frac{1}{2}\alpha^{2}x_{2}\partial_{x_{1}%
x_{1}}+\alpha\sigma\rho x_{2}\partial_{x_{1}x_{2}}+\frac{1}{2}\sigma^{2}%
x_{2}\partial_{x_{2}x_{2}}.
\]
It thus follows with $f_{u}(x)=e^{\mathfrak{i}u^{\top}x}$ that%
\[
\frac{Af_{u}(x)}{f_{u}(x)}=-\frac{1}{2}\alpha^{2}x_{2}\mathfrak{i}u_{1}%
+\kappa\left(  \theta-x_{2}\right)  \mathfrak{i}u_{2}-\frac{1}{2}\alpha
^{2}x_{2}u_{1}^{2}-\sigma\alpha\rho x_{2}u_{1}u_{2}-\frac{1}{2}\sigma^{2}%
x_{2}u_{2}^{2}%
\]
with first order derivatives w.r.t. $u,$
\begin{align*}
\partial_{u_{1}}\frac{Af_{u}(x)}{f_{u}(x)}  &  =-\frac{1}{2}\alpha^{2}%
x_{2}\mathfrak{i}-\alpha^{2}x_{2}u_{1}-\alpha\sigma\rho x_{2}u_{2},\\
\partial_{u_{2}}\frac{Af_{u}(x)}{f_{u}(x)}  &  =\kappa\left(  \theta
-x_{2}\right)  \mathfrak{i}-\alpha\sigma\rho x_{2}u_{1}-\sigma^{2}x_{2}u_{2}.
\end{align*}
For the second derivatives we get%
\[
\partial_{u_{1}u_{1}}\frac{Af_{u}(x)}{f_{u}(x)}=-\alpha^{2}x_{2},\text{
\ \ }\partial_{u_{2}u_{2}}\frac{Ae^{\mathfrak{i}u^{\top}x}}{e^{\mathfrak{i}%
u^{\top}x}}=-\sigma^{2}x_{2},\text{ \ \ }\partial_{u_{1}u_{2}}\frac
{Ae^{\mathfrak{i}u^{\top}x}}{e^{\mathfrak{i}u^{\top}x}}=-\alpha\sigma\rho
x_{2},
\]
and the third order ones vanish. Thus, in multi-index notation we have by
(\ref{sym}) for $\left\vert \beta\right\vert $ $=0,$%
\[
\mathfrak{b}_{0}(x,u)=\kappa\theta\mathfrak{i}u_{2}+x_{2}\left(  -\frac{1}%
{2}\alpha^{2}\mathfrak{i}u_{1}-\kappa\mathfrak{i}u_{2}-\frac{1}{2}\alpha
^{2}u_{1}^{2}-\alpha\sigma\rho u_{1}u_{2}-\frac{1}{2}\sigma^{2}u_{2}%
^{2}\right)  ,
\]
whence%
\begin{align*}
\mathfrak{b}_{0}^{0}(u)  &  =\kappa\theta\mathfrak{i}u_{2},\\
\mathfrak{b}_{0,e_{1}}^{1}(u)  &  =0,\text{ \ \ }\mathfrak{b}_{0,e_{2}}%
^{1}(u)=-\frac{1}{2}\alpha^{2}\mathfrak{i}u_{1}-\kappa\mathfrak{i}u_{2}%
-\frac{1}{2}\alpha^{2}u_{1}^{2}-\alpha\sigma\rho u_{1}u_{2}-\frac{1}{2}%
\sigma^{2}u_{2}^{2}%
\end{align*}
with $e_{1}:=(1,0),$ $e_{2}:=(0,1).$ For $\left\vert \beta\right\vert $ $=1,$
(\ref{sym}) yields
\begin{align*}
\mathfrak{b}_{(1,0)}(x,u)  &  =\allowbreak-\frac{1}{2}\alpha^{2}x_{2}%
+\alpha^{2}x_{2}u_{1}\mathfrak{i}+\alpha\sigma\rho x_{2}u_{2}\mathfrak{i,}\\
\mathfrak{b}_{(0,1)}(x,u)  &  =\kappa\left(  \theta-x_{2}\right)
+\alpha\sigma\rho x_{2}u_{1}\mathfrak{i}+\sigma^{2}x_{2}u_{2}\mathfrak{i,}%
\end{align*}
whence%
\[
\mathfrak{b}_{(1,0)}^{0}(u)=\mathfrak{b}_{(1,0),e_{1}}^{1}(u)=0,\text{
\ \ }\mathfrak{b}_{(1,0),e_{2}}^{1}(u)=-\frac{1}{2}\alpha^{2}+\alpha^{2}%
u_{1}\mathfrak{i}+\alpha\sigma\rho u_{2}\mathfrak{i}%
\]
and%
\[
\mathfrak{b}_{(0,1)}^{0}(u)=\kappa\theta,\text{ \ \ }\mathfrak{b}%
_{(0,1),e_{1}}^{1}(u)=0,\text{ \ \ }\mathfrak{b}_{(0,1),e_{2}}^{1}%
(u)=-\kappa+\alpha\sigma\rho u_{1}\mathfrak{i}+\sigma^{2}u_{2}\mathfrak{i}%
\]
Next, for $\left\vert \beta\right\vert $ $=2,$ (\ref{sym}) yields
\[
\mathfrak{b}_{(2,0)}(x,u)=\alpha^{2}x_{2},\text{ \ \ }\mathfrak{b}%
_{(0,2)}(x,u)=\sigma^{2}x_{2},\text{ \ \ }\mathfrak{b}_{(1,1)}(x,u)=\alpha
\sigma\rho x_{2},
\]
whence%
\begin{align*}
\mathfrak{b}_{(2,0)}^{0}(u)  &  =\mathfrak{b}_{(2,0),e_{1}}^{1}(u)=0,\text{
\ \ \ }\mathfrak{b}_{(2,0),e_{2}}^{1}(u)=\alpha^{2},\\
\mathfrak{b}_{(0,2)}^{0}(u)  &  =\mathfrak{b}_{(0,2),e_{1}}^{1}(u)=0,\text{
\ \ }\mathfrak{b}_{(0,2),e_{2}}^{1}(u)=\sigma^{2},\\
\mathfrak{b}_{(1,1)}^{0}(u)  &  =\mathfrak{b}_{(1,1),e_{1}}^{1}(u)=0,\text{
\ \ }\mathfrak{b}_{(1,1),e_{2}}^{1}(u)=\alpha\sigma\rho,
\end{align*}
and for $\left\vert \beta\right\vert \geq3,$ we trivially find%
\[
\mathfrak{b}_{\beta}(x,u)=0.
\]

\section{Generator and $\mathfrak{b}_{\beta}$ for the HSDJ model}

\label{AppB} By conferring (\ref{Gen}), (\ref{ItoLev}), and (\ref{HSDJ}), we
have in fact%
\[
v(x,dz)=v^{0}(dz)+x^{\top}v^{1}(dz)=\lambda_{0}\mu_{0}(z_{1})\delta_{0}%
(z_{2})dz_{1}dz_{2}+x_{2}\lambda_{1}\mu_{1}(z_{1})\delta_{0}(z_{2}%
)dz_{1}dz_{2}%
\]
with $\delta_{0}$ being the Dirac delta function, that is the (singular)
density of the Dirac probability measure $\mathbb{R}$ concentrated in
$\left\{  0\right\}  .$ Thus, the generator of the HSDJ model is given by
\begin{align*}
Af\,(x_{1},x_{2})  &  =\left(  -\lambda_{0}a_{0}-\left(  \frac{1}{2}\alpha
^{2}+\lambda_{1}a_{1}\right)  x_{2}\right)  \partial_{x_{1}}f+\kappa\left(
\theta-x_{2}\right)  \partial_{x_{2}}f\\
&  +\frac{1}{2}\alpha^{2}x_{2}\partial_{x_{1}x_{1}}f+\alpha\sigma\rho
x_{2}\partial_{x_{1}x_{2}}f+\frac{1}{2}\sigma^{2}x_{2}\partial_{x_{2}x_{2}}f\\
&  \!\!\!\!\!+\int_{\mathbb{R}}\left[  f(x_{1}+z_{1},x_{2})-f(x_{1}%
,x_{2})-z_{1}\partial_{x_{1}}f\right]  \left(  \lambda_{0}\mu_{0}(z_{1}%
)dz_{1}+x_{2}\lambda_{1}\mu_{1}(z_{1})dz_{1}\right)  .
\end{align*}
Since we are dealing with jump probability densities rather than infinite jump
measures, as in the case of infinite activity processes, the generator may be
written as%
\begin{gather*}
Af\,(x_{1},x_{2})=\left(  -\lambda_{0}\left(  \mathfrak{m}_{0}+a_{0}\right)
-\left(  \frac{1}{2}\alpha^{2}+\lambda_{1}\left(  \mathfrak{m}_{1}%
+a_{1}\right)  \right)  x_{2}\right)  \partial_{x_{1}}f\\
+\kappa\left(  \theta-x_{2}\right)  \partial_{x_{2}}f+\frac{1}{2}\alpha
^{2}x_{2}\partial_{x_{1}x_{1}}f+\alpha\sigma\rho x_{2}\partial_{x_{1}x_{2}%
}f+\frac{1}{2}\sigma^{2}x_{2}\partial_{x_{2}x_{2}}f\\
\!\!\!\!\!+\lambda_{0}\int_{\mathbb{R}}\left[  f(x_{1}+y,x_{2})-f(x_{1}%
,x_{2})\right]  \mu_{0}(y)dy\\
\!\!\!\!\!+x_{2}\lambda_{1}\int_{\mathbb{R}}\left[  f(x_{1}+y,x_{2}%
)-f(x_{1},x_{2})\right]  \mu_{1}(y)dy,
\end{gather*}
using (\ref{mom}).

With $f_{u}(x)=e^{\mathfrak{i}u^{\top}x}$ we so obtain,%
\begin{gather*}
\frac{Af_{u}(x)}{f_{u}(x)}=\left(  -\lambda_{0}\left(  \mathfrak{m}_{0}%
+a_{0}\right)  -\left(  \frac{1}{2}\alpha^{2}+\lambda_{1}\left(
\mathfrak{m}_{1}+a_{1}\right)  \right)  x_{2}\right)  \mathfrak{i}u_{1}\\
+\kappa\left(  \theta-x_{2}\right)  \mathfrak{i}u_{2}-\frac{1}{2}\alpha
^{2}x_{2}u_{1}^{2}-\alpha\sigma\rho x_{2}u_{1}u_{2}-\frac{1}{2}\sigma^{2}%
x_{2}u_{2}^{2}\\
+\lambda_{0}\psi_{0}(u_{1})+x_{2}\lambda_{1}\psi_{1}(u_{1})
\end{gather*}
with%
\begin{equation}
\psi_{i}(\xi):=\int_{\mathbb{R}}\left(  e^{\mathfrak{i}\xi y}-1\right)
\mu_{i}(y)dy,\text{ \ \ }i=0,1. \label{defp}%
\end{equation}
Note that we have%
\begin{equation}
\mathfrak{m}_{i}+a_{i}=\psi_{i}(-\mathfrak{i}),\text{ \ \ }i=0,1.
\label{defp1}%
\end{equation}
The first order derivatives w.r.t. $u$ are,
\begin{align*}
\partial_{u_{1}}\frac{Af_{u}(x)}{f_{u}(x)}  &  =-\lambda_{0}\left(
\mathfrak{m}_{0}+a_{0}\right)  \mathfrak{i}-\left(  \frac{1}{2}\alpha
^{2}+\lambda_{1}\left(  \mathfrak{m}_{1}+a_{1}\right)  \right)  \mathfrak{i}%
x_{2}\\
&  -\alpha^{2}x_{2}u_{1}-\alpha\sigma\rho x_{2}u_{2}+\lambda_{0}%
\partial_{u_{1}}\psi_{0}(u_{1})+x_{2}\lambda_{1}\partial_{u_{1}}\psi_{1}%
(u_{1})\\
\partial_{u_{2}}\frac{Af_{u}(x)}{f_{u}(x)}  &  =\kappa\left(  \theta
-x_{2}\right)  \mathfrak{i}-\alpha\sigma\rho x_{2}u_{1}-\sigma^{2}x_{2}u_{2}.
\end{align*}
For the second order derivatives we have%
\begin{align*}
\partial_{u_{1}u_{1}}\frac{Af_{u}(x)}{f_{u}(x)}  &  =-\alpha^{2}x_{2}%
+\lambda_{0}\partial_{u_{1}u_{1}}\psi_{0}(u_{1})+x_{2}\lambda_{1}%
\partial_{u_{1}u_{1}}\psi_{1}(u_{1})\\
\partial_{u_{1}u_{2}}\frac{Af_{u}(x)}{f_{u}(x)}  &  =-\alpha\sigma\rho
x_{2},\text{ \ \ }\partial_{u_{2}u_{2}}\frac{Af_{u}(x)}{f_{u}(x)}=-\sigma
^{2}x_{2},
\end{align*}
and for multi-indices $\beta$ with $\left\vert \beta\right\vert \geq3,$ i.e.
the higher order ones,%
\[
\partial_{u^{\beta}}\frac{Af_{u}(x)}{f_{u}(x)}=\left\{
\begin{tabular}
[c]{l}%
$\lambda_{0}\partial_{u_{1}^{\left\vert \beta\right\vert }}\psi_{0}%
(u_{1})+x_{2}\lambda_{1}\partial_{u_{1}^{\left\vert \beta\right\vert }}%
\psi_{1}(u_{1})$ \ \ for \ \ $\beta=(\left\vert \beta\right\vert ,0),$\\
$0$ \ \ if \ \ $\beta\neq(\left\vert \beta\right\vert ,0).$%
\end{tabular}
\ \ \right.
\]
Hence the ingredients (\ref{sym}) of the recursion (\ref{Rec}) are in
multi-index notation as follows.

$\left\vert \beta\right\vert =0:$%
\begin{gather*}
\mathfrak{b}_{0}(x,u)=-\lambda_{0}\left(  \mathfrak{m}_{0}+a_{0}\right)
\mathfrak{i}u_{1}+\kappa\theta\mathfrak{i}u_{2}+\lambda_{0}\psi_{0}(u_{1})\\
+x_{2}\left(  \lambda_{1}\psi_{1}(u_{1})-\left(  \frac{1}{2}\alpha^{2}%
+\lambda_{1}\left(  \mathfrak{m}_{1}+a_{1}\right)  \right)  \mathfrak{i}%
u_{1}-\kappa\mathfrak{i}u_{2}-\frac{1}{2}\alpha^{2}u_{1}^{2}-\alpha\sigma\rho
u_{1}u_{2}-\frac{1}{2}\sigma^{2}u_{2}^{2}\right)
\end{gather*}
whence%
\begin{align*}
\mathfrak{b}_{0}^{0}(u)  &  =-\lambda_{0}\left(  \mathfrak{m}_{0}%
+a_{0}\right)  \mathfrak{i}u_{1}+\kappa\theta\mathfrak{i}u_{2}+\lambda_{0}%
\psi_{0}(u_{1}),\\
\mathfrak{b}_{0,e_{1}}^{1}(u)  &  =0,\text{ \ \ }\mathfrak{b}_{0,e_{2}}%
^{1}(u)=\lambda_{1}\psi_{1}(u_{1})-\left(  \frac{1}{2}\alpha^{2}+\lambda
_{1}\left(  \mathfrak{m}_{1}+a_{1}\right)  \right)  \mathfrak{i}u_{1}\\
&  -\kappa\mathfrak{i}u_{2}-\frac{1}{2}\alpha^{2}u_{1}^{2}-\alpha\sigma\rho
u_{1}u_{2}-\frac{1}{2}\sigma^{2}u_{2}^{2}.
\end{align*}
For $\left\vert \beta\right\vert $ $=1,$ (\ref{sym}) yields%
\begin{align*}
\mathfrak{b}_{(1,0)}(x,u)  &  =-\lambda_{0}\left(  \mathfrak{m}_{0}%
+a_{0}\right)  -\lambda_{0}\partial_{u_{1}}\psi_{0}(u_{1})\mathfrak{i}-\left(
\frac{1}{2}\alpha^{2}+\lambda_{1}\left(  \mathfrak{m}_{1}+a_{1}\right)
\right)  x_{2}\\
&  +\alpha^{2}x_{2}u_{1}\mathfrak{i}+\alpha\sigma\rho x_{2}u_{2}%
\mathfrak{i}-x_{2}\lambda_{1}\partial_{u_{1}}\psi_{1}(u_{1})\mathfrak{i}\\
\mathfrak{b}_{(0,1)}(x,u)  &  =\kappa\left(  \theta-x_{2}\right)
+\alpha\sigma\rho x_{2}u_{1}\mathfrak{i}+\sigma^{2}x_{2}u_{2}\mathfrak{i,}%
\end{align*}
whence%
\begin{align*}
\mathfrak{b}_{(1,0)}^{0}(u)  &  =-\lambda_{0}\left(  \mathfrak{m}_{0}%
+a_{0}\right)  -\lambda_{0}\partial_{u_{1}}\psi_{0}(u_{1})\mathfrak{i,}\text{
\ \ }\mathfrak{b}_{(1,0),e_{1}}^{1}(u)=0,\\
\mathfrak{b}_{(1,0),e_{2}}^{1}(u)  &  =-\left(  \frac{1}{2}\alpha^{2}%
+\lambda_{1}\left(  \mathfrak{m}_{1}+a_{1}\right)  \right)  +\alpha^{2}%
u_{1}\mathfrak{i}+\alpha\sigma\rho u_{2}\mathfrak{i}-\lambda_{1}%
\partial_{u_{1}}\psi_{1}(u_{1})\mathfrak{i}%
\end{align*}
and%
\begin{align*}
\mathfrak{b}_{(0,1)}^{0}(u)  &  =\kappa\theta,\text{ \ \ }\mathfrak{b}%
_{(0,1),e_{1}}^{1}(u)=0,\\
\mathfrak{b}_{(0,1),e_{2}}^{1}(u)  &  =-\kappa+\alpha\sigma\rho u_{1}%
\mathfrak{i}+\sigma^{2}u_{2}\mathfrak{i.}%
\end{align*}
Next, for $\left\vert \beta\right\vert $ $=2,$ (\ref{sym}) yields%
\begin{align*}
\mathfrak{b}_{(2,0)}(x,u)  &  =\alpha^{2}x_{2}-\lambda_{0}\partial_{u_{1}%
u_{1}}\psi_{0}(u_{1})-x_{2}\lambda_{1}\partial_{u_{1}u_{1}}\psi_{1}(u_{1}),\\
\mathfrak{b}_{(1,1)}(x,u)  &  =\alpha\sigma\rho x_{2},\\
\text{\ }\mathfrak{b}_{(0,2)}(x,u)  &  =\sigma^{2}x_{2}%
\end{align*}
whence%
\begin{align*}
\mathfrak{b}_{(2,0)}^{0}(u)  &  =-\lambda_{0}\partial_{u_{1}u_{1}}\psi
_{0}(u_{1}),\text{ \ \ }\mathfrak{b}_{(2,0),e_{1}}^{1}(u)=0,\\
\mathfrak{b}_{(2,0),e_{2}}^{1}(u)  &  =\alpha^{2}-\lambda_{1}\partial
_{u_{1}u_{1}}\psi_{1}(u_{1}),\\
\mathfrak{b}_{(1,1)}^{0}(u)  &  =\mathfrak{b}_{(1,1),e_{1}}^{1}(u)=0,\text{
\ \ }\mathfrak{b}_{(1,1),e_{2}}^{1}(u)=\alpha\sigma\rho,\\
\mathfrak{b}_{(0,2)}^{0}(u)  &  =\mathfrak{b}_{(0,2),e_{1}}^{1}(u)=0,\text{
\ \ }\mathfrak{b}_{(0,2),e_{2}}^{1}(u)=\sigma^{2}.
\end{align*}
For multi-indices $\beta$ with $\left\vert \beta\right\vert \geq3$ we get%
\[
\mathfrak{b}_{\beta}(x,u)=\left\{
\begin{tabular}
[c]{l}%
$\lambda_{0}\mathfrak{i}^{-\left\vert \beta\right\vert }\partial
_{u_{1}^{\left\vert \beta\right\vert }}\psi_{0}(u_{1})+x_{2}\lambda
_{1}\mathfrak{i}^{-\left\vert \beta\right\vert }\partial_{u_{1}^{\left\vert
\beta\right\vert }}\psi_{1}(u_{1})$ \ \ for \ \ $\beta=(\left\vert
\beta\right\vert ,0),$\\
$0$ \ \ if \ \ $\beta\neq(\left\vert \beta\right\vert ,0),$%
\end{tabular}
\ \right.
\]
whence%
\[
\mathfrak{b}_{\beta}^{0}(u)=\left\{
\begin{tabular}
[c]{l}%
$\lambda_{0}\mathfrak{i}^{-\left\vert \beta\right\vert }\partial
_{u_{1}^{\left\vert \beta\right\vert }}\psi_{0}(u_{1})$ \ \ for \ \ $\beta
=(\left\vert \beta\right\vert ,0),$\\
$0$ \ \ if \ \ $\beta\neq(\left\vert \beta\right\vert ,0),$%
\end{tabular}
\ \right.
\]
and%
\begin{align*}
\mathfrak{b}_{\beta,e_{1}}^{1}(u)  &  =0,\\
\mathfrak{b}_{\beta,e_{2}}^{1}(u)  &  =\left\{
\begin{tabular}
[c]{l}%
$\lambda_{1}\mathfrak{i}^{-\left\vert \beta\right\vert }\partial
_{u_{1}^{\left\vert \beta\right\vert }}\psi_{1}(u_{1})$ \ \ for \ \ $\beta
=(\left\vert \beta\right\vert ,0),$\\
$0$ \ \ if \ \ $\beta\neq(\left\vert \beta\right\vert ,0).$%
\end{tabular}
\ \right.
\end{align*}

\bibliographystyle{plain}
\bibliography{HolTApplRef}

\end{document}